# HuBMAP Data Portal: a resource for multimodal spatial and single-cell data of healthy human tissues

**Authors**: Morgan L. Turner[1]*, Thomas C. Smits[1], Tiffany S. Liaw[1], Brendan Honick[2], Bill Shirey[3], Lisa Choy[1], Nikolay Akhmetov[1], Shaokun An[4], David Betancur[2], Dominic Bordelon[2], Karl Burke[3], Ivan Cao-Berg[2], John Conroy[1], Chris Csonka[2], Penny Cuda[5], Sean Donahue[5], Stephen A Fisher[6], Derek Furst[3], Ed Hanna[2], Josef Hardi[7], Tabassum Kakar[1], Mark S. Keller[1], Devin Lange[1], Xiang Li[2], Yan Ma[1], Alison McWilliams[2], Austen Money[1], Richard Morgan[3], Eric Mörth[1], Juan Muerto[2], Mark A. Musen[7], Emily Nic[2], Martin J. O'Connor[7], Gesina Phillips[2], Alexander J. Ropelewski[2], Ryan Sablosky[2], Sravani Saripalli[4], Max Sibilla[3], Derek Simmel[2], Alan Simmons[3], Xu Tang[4], Joel Welling[2], Zhou Yuan[3], Martin Hemberg[4], HuBMAP Consortium[+], Matthew Ruffalo[5], Jonathan Silverstein[3], Philip Blood[2], and Nils Gehlenborg[1]*

*Corresponding authors:
Morgan L. Turner, morgan_turner@hms.harvard.edu
Nils Gehlenborg, nils@hms.harvard.edu

[1] Harvard Medical School, Boston, MA, USA 02115

[2] Pittsburgh Supercomputing Center, Carnegie Mellon University, Pittsburgh, PA, USA 15213

[3] University of Pittsburgh, Pittsburgh PA, USA 15260

[4] Gene Lay Institute of Immunology and Inflammation, Brigham and Women's Hospital, Harvard Medical School and Massachusetts General Hospital, Boston, MA, USA 02115

[5] Carnegie Mellon University, Pittsburgh, PA, USA 15213

[6] University of Pennsylvania, Philadelphia, PA, USA 19104

[7] Stanford University, Stanford, CA, USA 94305

[+] A list of consortium authors and their affiliations appears at the end of the paper

## Abstract

The NIH Human BioMolecular Atlas Program (HuBMAP) Data Portal (https://portal.hubmapconsortium.org/) serves as a comprehensive repository for multimodal, multi-scale spatial and single-cell data from healthy human tissues. As of August 2026, the portal hosts 9,316 public datasets from 26 data types spanning 29 organ classes across 501 donors. Portal infrastructure and user interfaces support data search and discovery, visualization, and analysis directly in web browsers. These capabilities include metadata- and data-driven search, collaborative Workspaces with access to high-performance compute, and interactive Vitessce visualizations across non-spatial, 2D, and 3D spatial datasets. Data-type-specific uniform processing pipelines and rigorous quality control processes ensure comparability of results across laboratories, organs, and donors, while externally processed community-contributed datasets provide complementary perspectives. Here we describe portal functionality, infrastructure, and design, and highlight its role as a platform for large-scale spatial single-cell research across diverse data types, organs, and scales.

## Introduction

The Human BioMolecular Atlas Program (HuBMAP) is an NIH-sponsored program (https://commonfund.nih.gov/hubmap) that started in 2018 with the goal of creating a widely accessible, spatially-resolved, multi-scale reference atlas of healthy human organs and tissues at single-cell resolution[1]. Throughout the 8-year funded effort, over 600 HuBMAP Consortium members from more than 60 institutions have collaborated towards this goal, harnessing recent technological advances in molecular characterization of cell types and spatial mapping of complex human tissues[2]. Drawing upon biological and computational expertise within and beyond the consortium, the HuBMAP Data Portal (https://portal.hubmapconsortium.org/) was developed as a primary resource for open access to experimental tissue data and reference atlas data generated by HuBMAP, serving both consortium members and the broader research community.

The HuBMAP Consortium is organized into three components: Tissue Mapping Centers (TMC); innovative technologies groups (transformative technology development (TTD) and rapid technology implementation (RTI)); and the HuBMAP Integration, Visualization, and Engagement (HIVE). TMCs have generated data across a wide range of organs and biological variables such as age, sex, and race, selecting experimental technologies based on suitability for tissue types and research questions. Achieving large-scale cross-organ analyses, however, required overcoming two persistent challenges: inconsistent data outputs across generation methods, and variability in pipelines over time. To meet these challenges, the portal developed standardized schemas and data-type-specific uniform processing pipelines to harmonize outputs across modalities. Such harmonization efforts enabled analytical and visualization tools to be constructed and interoperate—enabling integrated analyses across data types, organs, and scales that would otherwise be difficult to achieve.

The initial consortium planning period (2018-2022) focused on creating ontologies, protocols, and pipelines necessary to construct a reference atlas for the human body, preparing teams for data generation and expanding infrastructure during the production phase (2022-2026)[2]. The multi-institutional HIVE team led the planning and development of the portal, informed by expertise and needs from the consortium and broader spatial biology community. Several requirements were established, including: 1) User Audiences: The portal must serve a wide range of users, needs, and workflows, particularly for data curation, search, analysis, visualization, and download. 2) Scope of Data: The portal must primarily host data from healthy human tissues. 3) Tools: The portal must provide tools supporting spatial data from multiple individuals across anatomical scales (organ to subcellular) following findable, accessible, interoperable, and reusable (FAIR)[3] data principles. 4) Standardization: The portal must standardize data and infrastructure to harmonize outputs across technologies[2], establish community metadata standards with subject matter experts and algorithm developers[4,5]; and develop integrated analytical and visualization tools at scale. 5) Sustainability: The portal must ensure long-term availability of open tools, data, and infrastructure to support research beyond the funding period. These requirements informed the design of the portal user interface (UI) and user experience (UX). We followed a user-centric design approach to develop query-driven interfaces that support exploration of complex biological human data (see Methods for detail on user-driven portal design)[6].

The HuBMAP Data Portal offers several capabilities that distinguish it from other data portals. From a data ingestion perspective, the portal supports hybrid connections between both uniformly processed data and Externally Processed Integrated Collections (EPICs), unlike most data portals which typically contain either one or the other. HuBMAP serves as a spatially resolved, multimodal complement to bulk-tissue resources such as GTEx[7] and transcriptomic platforms such as CZ CELLxGENE Discover[8], and broader atlas initiatives such as the Human Cell Atlas[9] and the Human Tumor Atlas Network (HTAN)[10,11]. By integrating spatial, single-cell, and multimodal data across diverse tissues and donors within a single portal, HuBMAP provides data at a scope and scale not currently available within any single existing resource. Few portals integrate raw data, processed data, and user-provided analysis results across a large breadth of data types within a single interface, and this approach has helped drive improvements in data format standardization. From an analysis perspective, integrated Vitessce visualizations[12] facilitate interactive data exploration directly on datasets within the portal, allowing users to quickly engage with data before selecting for further analysis or download. Workspaces enable lightweight exploration of public HuBMAP data and user-provided data without requiring local downloads, with templates serving as instructional guides for programmatically interacting with HuBMAP data using HuBMAP tools and APIs. Users can also explore data in Python or R through a HuBMAP-hosted JupyterLab environment. From a standardization perspective, the portal reflects a HuBMAP-wide effort spanning a large number of organs and data types[4]. The resulting standardized nomenclature, metadata, and cross-assay alignment enable large-scale integrated analyses across data types, tissue types, and anatomical scales (whole organ to subcellular)[2]. This standardization is achieved in part through the portal's integration with the Human Reference Atlas (HRA)[5] and Unified Biomedical Knowledge Graph (UBKG)[13], which link ontologies and biomedical community standards with HuBMAP portal data.

Here, we introduce the HuBMAP Data Portal as a FAIR-compliant resource for spatially resolved, multimodal single-cell data from healthy human tissues across diverse data types, organs, and scales (Fig. 1). We describe portal functionality and design across data submission and processing, search and discovery, visualization, and analysis, and highlight its role as a biological baseline against which disease can be understood and measured.

# Results

## HuBMAP Data on the Portal

The HuBMAP Data Portal contains multi-scale healthy human tissue data at single-cell and near-single-cell resolution, and hosts 9,316 published datasets (as of August 2026) from 26 data types spanning 29 organ classes across 501 donors (Fig. 2a). Donors (Fig. 2b,c) include 284 females and 209 males. Donor ages (Fig. 2b) range from 0 to 90 years (donors over 89 are set to an age of 90, in compliance with HIPAA regulations). Donor races (Fig. 2c) primarily consist of White, Black or African American, and Asian individuals, with relatively low sampling from underrepresented groups. Donor composition reflects tissue availability and research priorities of NIH-funded HuBMAP projects rather than a population-representative sampling strategy (see Methods for detail on donor metadata, including the use of "unknown"). Organ classes (Fig. 2d-g) are specified by the right or left side for bilateral pairings. Datasets (Fig. 2d) are highest in count from placenta, lung (left and right), uterus, kidney (left and right), and small and large intestine tissue. Among the 5,142 samples (Fig. 2e), skin, kidney, uterus, and placenta are the most common organ sources. Samples are categorized as organs, blocks, sections, or suspensions, with tissue sections (2,659) and blocks (1,660) representing the majority of spatial data. These reflect contiguous anatomical regions predominantly from whole tissue, consistent with HuBMAP's goal of spatially resolved single-cell mapping. There are 584 unique donor organs with data on the portal (Fig. 2f), which together have an average sample number per donor organ of 8.8 (Fig. 2g). Data types (Fig. 2h-i) span both spatial and non-spatial, as well as single and multimodal assays. By dataset count (Fig. 2h), the most common non-spatial data types include RNAseq (1,605 datasets), probe-based RNAseq (1,362 datasets), and ATACseq (1,125). The most common spatial data types include GeoMx (1,362), Histology (889), MIBI (429), Autofluorescence (356), and CODEX (255). The most rare data types include spatial methods such as 3D and 2D IMC (3, 13 respectively), Light Sheet (11), CosMx (12), MUSIC (14), and DESI (15). Cell type annotations have been produced for 165 uniformly processed sc/snRNAseq, 10x Multiome, and Slide-seq datasets across three organs currently supported (heart, lung, and kidney). Vitessce visualizations (Fig. 2i) are provided for 5,135 of the 9,316 portal datasets and are available for 19 of the 26 data types across the portal.

The data model on the portal follows the typical data generation sequence: *donor* > *sample* > *assay* > *raw dataset* > *processed dataset*. Some data types (e.g., SNARE-seq2, 10x Multiome) combine multiple raw assay datasets (e.g., RNAseq and ATACseq) into a single processed dataset, as shown in the Dataset Relationship Diagram and Provenance Graph (see below). Tissue from donor organs is sampled and then assayed using experimental technologies chosen by data generators. Raw data is obtained from an assay or sequencing instrument and may undergo additional transformation or interpretation prior to submission by data generators via the HuBMAP ingest portal (https://ingest.hubmapconsortium.org/). Once submitted, the transition from raw to processed datasets is facilitated by the execution of uniform processing pipelines within the portal infrastructure, or by an external analysis executed in the lab of a data submitter in the cases of EPICs (Fig. 1). Processing typically involves cleaning, normalizing, aligning, and summarizing raw data. The specific steps applied vary by data type, with bulk, single-cell, and spatial data types each undergoing distinct processing appropriate to their modality (see Methods for detail on data submission, ingestion, processing, and provenance).

Submitted data and metadata are curated through a sequence of ingestion, processing, and quality control before being published on the portal. Within the HuBMAP Consortium, the ultimate responsibility for data quality lies with the data submitter, who has domain expertise in the relevant technology and data type. Data submitters review processed datasets prior to publication, applying quality criteria beyond what is automated within the pipelines. Some variability in sample quality is an inherent possibility in a large multi-site consortium, where data generation centers each bring unique organ- and technology-specific expertise. The portal supports transparency by representing the complete processing history for each dataset, and users are encouraged to inspect datasets directly, including through available Vitessce visualizations, and apply their own quality criteria before downloading or further use.

Within this processing framework, several additional pipelines and tools contribute to how data are represented and made accessible on the portal. Images are processed via an image pyramid pipeline resulting in a "support" dataset that enables Vitessce visualizations on the portal. Cell type annotations are generated within the sc/snRNAseq pipeline for supported tissues (currently heart, lung, and kidney) by Azimuth[14], a reference-based single-cell annotation tool designed for transcriptomic data; a pan-human extension of Azimuth spanning organs beyond current tissue-specific references is under development[15]. In data visualizations and downloadable pipeline outputs, cell type labels are mapped to cell ontology terms to facilitate comparisons across datasets. Annotations are explorable within the Cell Population Plot tool *scellop*[16], Vitessce, Cell Type Pages, and Datasets Search. External analyses that fall outside of internal HuBMAP uniform processing pipelines are uploaded and indexed as EPICs, which are otherwise treated as standard portal datasets alongside raw and processed data throughout portal interfaces. For each dataset, a Dataset Relationship Diagram illustrates a unified view of raw data, any applied processing actions, and resulting processed data. Raw data protocols are available at protocols.io, and analysis pipelines and metadata schemas are available at https://github.com/hubmapconsortium. The resulting published datasets are available to consortium members and public consumers of HuBMAP data under a Creative Commons Attribution 4.0 International License (CC BY 4.0). The HuBMAP External Data Sharing Policy (https://hubmapconsortium.org/policies/external-data-sharing-policy/) provides additional guidance on appropriate use and attribution for consortium and public users.

## Data Types

The HuBMAP Data Portal supports and contains 26 data types across 29 organ classes (Fig. 3). For both spatial (2D & 3D) and non-spatial data, data types cover a rich spectrum of analytes, including RNA, protein, DNA, endogenous fluorophore, chromatin, peptide, polysaccharide, lipid, metabolite, and combinations thereof.

The portal hosts an extensive collection of 2D spatial data. CODEX (Akoya Biosciences) measures protein abundance at single-cell resolution using multiplexed immunofluorescence and has been applied to large and small intestine[17], spleen, lymph node, and thymus tissue. Its successor PhenoCycler, treated as a separate data type due to substantial differences in hardware, output data structure, imaging area, and biomarker capacity, has been applied to bone marrow, eye, fallopian tube, heart, kidney, lung, and lymph node tissue. Multiplexed Ion Beam Imaging (MIBI)[18], which measures proteins at single-cell resolution, has been applied to uterus[19] and bone marrow tissue. Cell DIVE (Leica Biosystems), a rarer 2D spatial data type, measures the abundance of multiple biomarkers at single-cell resolution within spatial context and has been applied to heart, placenta, and skin tissue[20,21]. Autofluorescence microscopy captures endogenous tissue fluorescence that can be used to integrate data from multiple modalities and align tissues within 3D experiments, and has been applied to bronchus, eye, kidney, lung, and pancreas tissue. Visium (no probes) measures whole-transcriptome gene expression and has been applied to ovary, uterus, and fallopian tube tissue, as well as to kidney tissue in patients with kidney stone disease[22]. Matrix-Assisted Laser Desorption Ionization (MALDI) imaging mass spectrometry[23] maps metabolites, lipids, and in some cases proteins, within an imaged anatomical context and has been applied to bone marrow and kidney tissue. The portal also hosts 2D IMC, CosMx, DESI, Histology, seqFish, and Slide-seq data (Fig. 3).

The portal also hosts 3D spatial data. Light sheet fluorescence microscopy, which optically sections and images tissue volumes at high resolution in conjunction with protein abundance measurements, has been applied to kidney[24], lymph node, spleen, and thymus tissue. 3D imaging mass cytometry (IMC) provides insights into cellular microenvironments and tissue architecture and has been applied to lymph node, spleen, and thymus tissue.

Multimodal spatial assays, which map multiple molecular types and spatial coordinates in the same dataset, are less common on the portal but have a significant scientific impact. Multinucleic acid interaction mapping in single cells (MUSIC) profiles multiplex chromatin interactions, gene expression and RNA–chromatin associations within individual nuclei, and has been applied to brain tissue[25]. The GeoMx Digital Spatial Profiler (NanoString) simultaneously measures RNA and protein abundance from selected regions of interest and has been applied to placenta tissue. Xenium (10x Genomics) simultaneously measures RNA transcripts and protein abundance at single-cell resolution using targeted in situ hybridization and antibody-based imaging, and has been applied to the small intestine; additional organs, including kidney[26], are currently in QA.

The portal hosts an extensive collection of non-spatial molecular profiling data. RNAseq, which encompasses bulk, single-cell, and single-nucleus RNA sequencing, measures RNA abundance and is applied to nearly all tissue types on the portal. Probe-based RNA sequencing, which uses targeted probes to capture and enrich specific RNA regions for in-depth analysis, has been applied to placenta tissue. ATACseq, which encompasses bulk, single cell, and single-nucleus ATAC sequencing, measures chromatin accessibility and is widely applied across organs. Each of these methods is well-represented in the portal, and they have been applied together for large studies on intestine[17] and kidney[27] tissue.

Less common non-spatial data types are also impactful. 10x Multiome, a non-spatial multimodal assay that measures RNA and chromatin accessibility at single-cell resolution, is applied to ovary, uterus, and fallopian tube tissue for female reproductive research, as well as bladder, bone marrow, kidney, and lymph node tissue. Another rare but impactful non-spatial data type is CyTOF, which measures up to 50 protein markers and has been applied to bone marrow and blood. Liquid chromatography-mass spectrometry (LC-MS) provides high-sensitivity profiling of proteins, metabolites, and lipids from tissue extracts and represents one of the most dataset-rich non-spatial data types on the portal, applied across heart, kidney, large and small intestine, lung, pancreas, placenta, and spleen tissue. The portal also hosts whole-genome sequencing (WGS) and SNARE-seq2 data, the latter applied to lung tissue[28] (Fig. 3).

## Searching and Discovering Datasets

Core to HuBMAP Data Portal functionality is the ability to search and discover data using different approaches (Fig. 4). To support users in accessing relevant datasets, the portal provides metadata-driven and data-driven search queries, curated data groupings, and direct programmatic API access. These different entry points enable users with a range of expertise and scientific goals to integrate HuBMAP data and portal functionality with preferred workflows.

For metadata-driven queries, the primary search interface is the Datasets Search page (https://portal.hubmapconsortium.org/search/datasets) (Fig. 4a-b), where users can search, explore, select, and download datasets matching specific metadata parameters. This approach is often employed by experimental biologists to identify healthy reference data for a specific disease (e.g., bladder cancer) or by computational biologists to identify specific datasets (e.g., raw datasets of breast tissue containing protein data) for training models in support of therapeutic development. The user workflow is built around faceted browsing in the Filter & Browse mode, where datasets can be filtered by metadata provided by the data contributor, including assay type, organ, analyte class, sample category (organ, block, section, suspension), and UBKG-referenced donor attributes (e.g., sex, age, race, BMI). Metadata generated from the HuBMAP uniform processing pipelines is also available, including raw and processed data states, HuBMAP or external lab processing type, single or multiple assay modalities, processed cell type annotations, dataset publication dates, visualization availability, and more (see Methods for detail on metadata). From search results, users can explore individual Dataset pages, manage datasets in Workspaces, or save datasets using the My Lists feature. For downloading, users can generate a manifest for bulk Globus transfer via the HuBMAP Command Line Transfer utility (CLT; https://docs.hubmapconsortium.org/clt/index.html), or download a TSV of donor and sample metadata.

The portal also provides metadata-driven Dataset Search using the Say & See mode, a natural language chat interface powered by YAC (Yet Another Chatbot)[29]. Say & See allows users to query metadata conversationally and receive interactive visualizations in response (Fig. 4c), lowering the barrier for users less familiar with faceted search interfaces. By default, Say & See queries a fast shared cache of HuBMAP's public index, with an option to switch to an authenticated view for querying non-public datasets accessible to the user's account. Similar to the results of Filter & Browse mode, results can be downloaded as a manifest file for bulk download of data from Globus or as CSVs of related donors, samples, and datasets metadata for further analysis.

For data-driven queries, the Biomarker & Cell Type Search (https://portal.hubmapconsortium.org/search/biomarkers-celltypes) (Fig. 4d-e) allows users to retrieve datasets based on the presence of a cell type, expression of a biomarker (transcriptomic, epigenomic, and proteomic abundance), or other statistical measures. Users choose from three query types: "Gene" for transcriptomic and epigenomic measurements, "Protein" for proteomic measurements, and "Cell Type" for cell type distribution. Two query methods are available to select from: *scFind* and *Cells Cross-Modality*. *scFind*[30] supports Gene and Cell Type queries across comprehensively indexed sc/snRNAseq (gene expression), ATACseq (chromatin accessibility), and Cell Type (prevalence) data. *Cells Cross-Modality* extends this to Protein queries and additionally covers proteomic measurements, making it the broader option for cross-modality gene, protein, and cell type searches. For gene queries, pathways from Reactome (https://reactome.org/)[31] can be used to efficiently select multiple genes simultaneously. Results include dataset lists with accompanying statistical summaries, interactive visualizations, and query-type-specific outputs such as cell type distribution plots and protein abundance charts.

The ability to query HuBMAP data using biologically-relevant parameters enhances its utility across a range of research contexts. Translational scientists can investigate baseline molecular and cellular profiles in healthy tissue to define normal activation profiles for comparison with disease states. Molecular or cell biologists can search biomarkers to determine which cell types within an organ express specified genes to provide context for functional studies. Finally, computational biologists can identify large uniformly processed datasets across multiple organs to train AI models or benchmark computational tools.

Beyond direct search, the portal offers biologically contextualized entry points through dedicated pages for Organs (https://portal.hubmapconsortium.org/organ), Biomarkers (https://portal.hubmapconsortium.org/biomarkers), and Cell Types (https://portal.hubmapconsortium.org/cell-types). The Organs page provides summary information and interactive graphs to filter datasets by assay type, donor metadata, and analyte class. Individual organ pages include an overview of organ function with links to UBERON ontologies[32], visualizations of spatially-registered tissue samples using the HRA Common Coordinate Framework[33], and a list of associated HuBMAP tissue samples. The Biomarkers page allows users to explore gene biomarkers and find detailed information on associated organs, cell types, and datasets, or search for a gene of interest by name or symbol. Individual gene pages provide an overview of gene function, links to external resources such as HUGO[34] and Ensembl[35], associated cell types, and datasets featuring the gene with an expression visualization. The Cell Types page provides browsable interactive graphs and a comprehensive list of cell types in HuBMAP samples identified by Azimuth. Each cell type page provides a detailed overview including description, distribution across organs, key marker genes, associated datasets, and interactive plots for visualizing and comparing cell type distributions.

Datasets are also organized thematically by domain experts through Publications (https://portal.hubmapconsortium.org/publications), Collections (https://portal.hubmapconsortium.org/collections), and Integrated Maps (https://portal.hubmapconsortium.org/integrated-maps). Publication pages pair peer-reviewed publications and preprints with referenced HuBMAP datasets and vignettes of relevant findings illustrated with interactive visualizations. Collections group HuBMAP datasets from related studies, and each is assigned a Digital Object Identifier (DOI) for citation and reference. Integrated Maps offer download-ready HuBMAP-wide consolidated data for datasets of a particular assay type and tissue, aggregated across multiple datasets.

Programmatic access to data is available for users who prefer computational methods. The portal is built upon a flexible hybrid cloud infrastructure, integrating applications, APIs, and resources supported by Amazon Web Services and Pittsburgh Supercomputing Center (see Börner et al. [5] for details on portal architecture and microservices). For example, the Search API (https://docs.hubmapconsortium.org/param-search/) offers parameterized search capabilities for simpler filtering and querying, while direct use of the Elasticsearch Domain Specific Language (DSL) (https://www.elastic.co/docs/explore-analyze/query-filter/languages/querydsl) provides greater flexibility for more complex queries.

## Visualizing Datasets

The HuBMAP Data Portal provides interactive visualization tools to enable users to explore diverse data types and biological questions directly in the portal without requiring downloads or software installation. Individual datasets are visualized using Vitessce[12], supporting 19 of the 26 data types and over half of all HuBMAP datasets (Fig. 2i; Fig. 5). Aggregate data can be visualized using the Cell Population Plot tool *scellop* on the organ page or in analysis Workspaces for cross-sample comparison (see below). In search interfaces, interactive and configurable graphs support high-level exploration of available HuBMAP data. Dataset provenance and processing relationships are also visualized.

Vitessce (https://vitessce.io/)[12] is a web-based framework for visualization of multimodal 2D and 3D spatial and single-cell data, embedded across the portal to provide visualizations of omics and imaging data alongside associated metadata. Visualizations have been tailored to available data and biological questions, providing users a consistent tool and interface across compatible HuBMAP data types (Fig. 5). Vitessce supports users across multiple stages of data analysis, providing interactive visualizations and programming libraries in Python, R, and JavaScript. The chart types supported by Vitessce—including dimensionality reduction scatterplots, spatial and imaging views, dot plots, and violin plots—reflect the visualization designs most commonly employed in single-cell atlas publications[36]. During HuBMAP data ingestion, users can inspect and conduct quality assurance on uniformly processed datasets directly in Vitessce. Multiple linked visualizations embedded directly on dataset pages (e.g., Fig. 4b, inset) allow users to explore published multimodal data. Users can launch a Jupyter Notebook instance to execute interactive computational analyses and embed interactive Vitessce visualizations through the corresponding R and Python packages (see below). For example, analysis packages such as SpatialQuery[37] integrate with Vitessce to display interactive outputs of identified cell type co-localization patterns. The underlying Vitessce visualization configuration can easily be shared for collaboration or included in analysis Workspaces, and is available for download. Visualizations and data files associated with HuBMAP research publications can be linked and made accessible on individual Publication Pages (https://portal.hubmapconsortium.org/publications). With access to standard consumer Virtual Reality hardware, volumetric data can be explored in an immersive 3D environment via Vitessce Link[38]. Vitessce Link also supports 3D Tissue Maps (Fig. 5, asterisk), a data format for multimodal 3D data comprising registered and aligned volumetric datasets that combine biomarker and point cloud data with segmentations of anatomical structures. 3D Tissue Maps enable interactive exploration of spatial relationships among cell types, functional tissue units (FTUs), and morphological structures on the portal (e.g., https://portal.hubmapconsortium.org/preview/3d-tissue-maps). Portal analytics reveal that engagement with Vitessce visualizations consistently accounts for over half of all user interactions, suggesting visual data exploration is a core use of the portal.

To facilitate cross-sample comparisons, we integrated *scellop* (https://github.com/hms-dbmi/scellop)[16], an interactive cell population plot viewer, into the portal (Fig. 6a,b). Using *scellop*, users can compare cell type counts and proportions within and across samples to discover patterns related to donor and sample information. *scellop* addresses limitations of traditional stacked bar chart designs, which rely on color encoding for cell types and face scalability challenges. Instead, *scellop* presents a multi-panel view with a central heatmap showing samples and cell types along its two axes. To examine specific samples, individual heatmap rows can be converted to (non-stacked) bar charts. This approach of combining chart types aids global pattern detection and comparisons of cell populations among samples. Data portal user studies informed key design features, including normalization options and grouping, sorting, or filtering by metadata (e.g., donor age). *scellop* automatically loads sample- and donor-level metadata, and can visualize any dataset containing cell type annotations (e.g., from Azimuth). *scellop* is available on Organ pages for organs with cell type annotations, currently Heart, Kidney, and Lung (e.g., http://portal.hubmapconsortium.org/organ/kidney), and available as a Workspace template (see below).

Beyond Vitessce and *scellop* visualizations, each dataset page includes diagrams illustrating dataset provenance and relationships to other datasets, samples, and donors (Fig. 6c,d). The Dataset Relationship Diagram illustrates which processing pipelines have been applied on the raw dataset. The Provenance Table and Graph show how the dataset is derived from a donor and tissue samples, and display relationships to other derived datasets. Together, these tools help users contextualize datasets within the portal.

## Analyzing Datasets

HuBMAP Workspaces (https://portal.hubmapconsortium.org/workspaces) are an integrated online analysis environment for Python and R, complementing the automated uniform processing pipelines and consortium-designed analysis workflows. Similar platforms include 4DN's JupyterHub (https://data.4dnucleome.org/tools/jupyterhub)[39] and the Biomedical Research Hub[40], both of which offer example analyses. The All of Us Research Workbench[41] and the Terra Workspaces in the AnVIL project (https://anvil.terra.bio) similarly support collaborative analysis environments.

Designed to bring user-driven analysis closer to the data, HuBMAP Workspaces support faster insights and democratize single-cell and spatial biology data analysis through four integrated capabilities. First, direct data access removes the need for local downloads, allowing users to inspect, analyze, and visualize data in place. Second, Workspaces are backed by computational resources at the Pittsburgh Supercomputing Center (PSC), with configurable options, including memory, number of CPUs, GPUs, time limits, and preinstalled packages. Third, fifteen analysis templates provide guided, assay-specific workflows demonstrating how HuBMAP tools and APIs can be applied. Each template includes a launchable example with sample data, and users can contribute their own. Finally, collaborative Workspaces allow users to invite others to receive a copy of the notebook files and datasets, streamlining shared analysis and workflow exchange.

Workspaces are deeply integrated into the portal and accessible through multiple entry points, as illustrated in Fig. 7. Users can create a Workspace linked to datasets and templates (e.g., analyzing spatial neighborhoods in Slide-seq data using Squidpy), launching a JupyterLab environment. Users can edit files, add or remove datasets, add new templates for additional analysis types, adjust launch configurations to obtain appropriate compute resources, and export figures and notebooks. Users can also run multiple concurrent Workspaces and share them with other HuBMAP Workspaces users.

A key feature that distinguishes HuBMAP Workspaces and increases data FAIRness is direct access to powerful compute resources co-located with the data at PSC, made available at no cost through the HuBMAP flexible hybrid cloud infrastructure. Users can access Workspaces by logging into the portal using Globus[42,43] and following the prompts to connect their institutional credentials. Compute capabilities are currently provided through PSC resources dedicated to HuBMAP, but as demand grows, PSC can expand capacity through HuBMAP-specific resource allocations on NSF ACCESS systems[44] located at PSC. These resources are immediately available to HuBMAP Consortium members by default, and available to non-consortium members on request by following sign-up instructions on the Workspaces page (https://hubmapconsortium.org/workspaces-sign-up/).

### Retrieving and Downloading Datasets

The HuBMAP Data Portal provides multiple methods for retrieving and downloading datasets, ranging from direct file downloads to bulk programmatic access. 1) Datasets selected through portal interfaces can be downloaded as a group. 2) Individual files from processed datasets can be downloaded directly through the file browser on each dataset page. 3) Using the HuBMAP CLT Globus[42,43] integration, users can generate a manifest file through portal search and use it to bulk download files from multiple Globus collections. 4) Users can navigate directly to the Globus HuBMAP endpoint to download data outside of the portal interface, which may be preferable for large transfers. 5) Raw sequencing data containing protected health information is hosted on the NIH database of Genotypes and Phenotypes (dbGaP), outside the portal, and can be downloaded directly following dbGaP access request approval. Access permissions on the portal are divided into protected and non-protected tiers, determining which datasets are publicly visible and which require additional authorization to download.

## Discussion

The HuBMAP Data Portal serves a global research community—from translational scientists defining healthy tissue baselines to computational biologists training AI models—as more than a data repository. Where most portals collect, process, and distribute consortium data, HuBMAP also enables rapid discovery and in-browser insight generation. Its combination of uniformly and externally processed data, raw and derived data types, and standardized cross-assay outputs within a single resource remains uncommon among biomedical data portals. This commitment to FAIR[3] data principles extends across portal tool development and is reflected in an evaluation of 28 biomedical databases, where the HuBMAP Data Portal was the only repository to satisfy all 29 FAIR assessment criteria[45].

The data-type-specific uniform processing pipelines and standardized schemas developed during the construction of the portal address core challenges in harmonizing outputs across data generation methods, organs, and metadata standards. Pipeline versioning ensures comparability across datasets processed at different points in time, allowing improvements without disrupting the broader collection. Long-running consortium efforts such as ENCODE[46] have demonstrated the sustained value of uniform processing pipelines and rigorous metadata standards for large-scale functional genomics data, establishing an important precedent for consortium-driven standardization at scale. The CEDAR metadata validation infrastructure employed by HuBMAP further supports standardization by enabling submitters to validate metadata prior to submission, reducing errors and improving consistency across the portal[4]. The diversity of the resulting data is unique among data repositories, enabling large-scale cross-organ integrative analyses. However, broader challenges remain. Evolving experimental technologies are producing increasingly large and complex datasets, and the lack of widely adopted community standards limits the scale of addressable research questions. Community standards exist for some data formats, such as OME-Zarr and OME-TIFF[47,48], but adoption remains inconsistent. Metadata standardization is a related challenge: despite growing recognition of its importance, most experiment types still lack community-endorsed metadata standards, making data sharing and reuse difficult[4]. The HuBMAP Consortium's coordinated approach, driven by community engagement and shared schemas, represents one of the more systematic efforts to address this challenge at scale.

Historically, the field has faced tension between uniform processing and the need for externally processed data to accommodate lab-specific analyses and research goals. With the growing need to perform large-scale analyses, uniform processing is increasingly necessary, as it ensures comparable results across a wide range of assays (e.g., bulk, single-cell, and multimodal assays). While adherence to standards often requires significant effort from data submitters, these efforts are essential for assembling a large-scale reference set. For many research groups, it remains beneficial to use, analyze, and publish findings from data processed outside of uniform pipelines[49]. The EPIC model of integrating externally processed data with uniformly processed data introduced in the portal satisfies both community needs. Data-generating labs can upload externally processed data, which can be integrated into downstream analyses alongside raw and uniformly processed data (Fig. 1). This experience also points to a broader lesson: earlier definition of common high-level data outputs, such as cell-by-gene matrices or image segmentations, enables a more flexible infrastructure adaptable for a wider range of community needs. Standardized outputs of this kind also have direct downstream value: they accelerate tool development and lower barriers to AI model training by providing predictable interfaces for extension and reuse. As data volumes and model complexity grow, consortia-driven adherence to community standards will likely be central to future advances in algorithm development and foundation model training.

While the portal supports a broad range of harmonized and standardized data, limitations exist. The scope of portal data is constrained by donor tissue availability and the research priorities of NIH-funded projects, resulting in a cohort that does not fully reflect human demographic diversity and remains sparse in spatial coverage and data type representation across organs (e.g., Fig. 2). Users should consider these limitations when employing HuBMAP data as a healthy reference or training set, as findings may not generalize across all demographic groups or tissue contexts. Spatial sampling at the organ level is uneven and biased by research questions, limiting the utility for some large-scale analyses. Cell type annotations, which can be queried to find datasets, visualized with *scellop* population plots, and analyzed in Workspace notebooks, are currently available for only three organs (heart, lung, and kidney). Expansion to additional organs will substantially increase the portal's utility for cell-type-driven analyses. Vitessce visualizations are currently available for 19 of the 26 data types, with coverage gaps reflecting the complexity of supporting emerging and rare data types such as CosMx and MUSIC. While advances in computational methods have increased data processing throughput, data generation remains limited by donor tissue availability and TMC processing capacity. Limited human resources constrain expansion of portal tools to operate on new or updated data formats, and scalability challenges will need to be addressed as data volumes grow and new experimental technologies emerge.

The HuBMAP Data Portal demonstrates how to collect, process, and communicate the data necessary to construct a human biomolecular reference atlas. Central to this effort, the portal has contributed to the construction of a 3D Human Reference Atlas (HRA)[5] of the healthy human body, providing spatially registered tissue data, data-type-specific uniform processing pipelines, cell type annotations, and visualization infrastructure through which atlas data are integrated, validated, and made accessible. The modular design of the portal supports long-term sustainability and has enabled direct reuse of components beyond HuBMAP. For example, the Common Fund's Cellular Senescence Network (SenNet) Program adopted several HuBMAP portal components in the construction of the SenNet portal (e.g., cloud-based hybrid infrastructure, metadata and directory schema, uniform processing pipelines, and Vitessce) for the study of senescent cells across human and other model organism tissues[50], accelerating SenNet portal development. Modular design also benefits end users directly, providing standardized metadata and data formats, interoperable tools, and consistent user interfaces across the portal's diverse datasets.

As disease atlases rapidly grow, the HuBMAP Data Portal's role as a healthy human reference provides a critical biological baseline against which disease can be understood, measured, and ultimately addressed. Realizing this across a broad research community requires design decisions that prioritize accessibility: lowering barriers through intuitive interfaces, natural language search, and analysis environments that bring compute directly to the data. The future of integrated biomedical research will rely on development and maintenance of large-scale data repositories such as the HuBMAP Data Portal, and large-scale coordination of teams and funding mechanisms will be required to sustain these efforts[51,52]. The HuBMAP Data Portal has been designed with this in mind by minimizing dependence on any single funding source or team, yet its long-term success will require sustained community investment.

# Methods

## HuBMAP Data Submission and Ingestion

To support the end goal of publishing datasets on the HuBMAP Data Portal, researchers at HuBMAP-funded components must follow a set of formalized procedures with the support of HIVE data curators (information science professionals who support open science through research data management practices). These workflows can be divided into two categories, data submission and data ingestion. Both are driven by a funded component's data submitter: a researcher, staff scientist, graduate student, post-doctoral researcher, or other professional designated to share their data with HuBMAP. Submitters work directly with HIVE data curators throughout this process, with submitters driving submission and curators pushing forward ingestion.

### *Data Submission*

Data submitters register bulk data uploads for one or more datasets of the same data type in the HuBMAP ingest portal (https://ingest.hubmapconsortium.org/). Registration generates a unique human-friendly HuBMAP ID along with a 32-character universally unique identifier (UUID) for the upload, which is then used to create a unique Globus-managed staging directory for the bulk data upload in the HIVE file system. This Globus-managed staging directory is kept "protected" by default in case human sequencing data is deposited. To prepare the bulk data upload for submission, a data submitter organizes the files from their experiment into one or more dataset directories on a local server running Globus that can then be uploaded to the unique Globus-managed staging directory on the HIVE file system via Globus Transfer.

These dataset directories, and the data files organized within them, must adhere to the directory schema for the data type being submitted. The submitter must also prepare an assay metadata file and a collaborators metadata file (tab separated values), each with a header and one row of values per dataset, and include this at the top level of the bulk upload directory. The HIVE maintains a centralized list of both directory and metadata schemas (https://docs.hubmapconsortium.org/metadata) for submitters' reference. Metadata schemas are created and published in the Center for Expanded Data Annotation and Retrieval (CEDAR)[53] and provided to end users as user-fillable spreadsheet templates. The HIVE has developed the CEDAR Metadata Spreadsheet Validator[54] that submitters can use to easily validate and quickly correct any errors in their metadata files prior to submission, to ensure accuracy and avoid wasted time going back and forth with HIVE data curators.

The data submission process also includes several additional steps that are independently carried out from the bulk data upload. These include preparing experimental protocols, donor metadata, tissue sample metadata, and spatial registration, which together significantly boost the FAIRness of HuBMAP data. These steps are detailed below.

Experimental protocols: The HIVE maintains a workspace on protocols.io (http://protocols.io) where funded components publish and mint DOIs for donor case selection, tissue sample preparation, and assay protocols relevant to their studies. DOIs for assay protocols are included in the assay metadata files, while the DOIs for the case selection and preparation protocols are included in the metadata for donors and tissue samples.

Donor metadata: HuBMAP does not use templates for whole human donor metadata. Clinical donor metadata is handled by the HIVE's Honest Broker Service, which abstracts out relevant descriptors from resources that submitters provide via a secure Health Insurance Portability and Accountability Act (HIPAA)-compliant Globus endpoint. Examples of such resources include spreadsheets, electronic medical records, PDFs, and REDCap studies. At minimum, the HIVE requires information about a donor's age, race, sex, and whether they were living or deceased at the time of sample collection. These generalized metadata values are used to programmatically populate titles for dataset DOI records as the HIVE publishes datasets. When one of those values is not present in the local metadata from a funded component, the HIVE uses a placeholder value of "Unknown" in the portal. In compliance with HIPAA regulations, donors over 89 are set to an age of 90. A full list of supported donor metadata documentation is available (https://docs.hubmapconsortium.org/donor.html).

Tissue sample metadata: HuBMAP specifies metadata templates and captures metadata for tissue blocks (https://hubmapconsortium.github.io/ingest-validation-tools/sample-block/current/), sections (https://hubmapconsortium.github.io/ingest-validation-tools/sample-section/current/),

and suspensions (https://hubmapconsortium.github.io/ingest-validation-tools/sample-suspension/current/). Donors, biological samples, and associated datasets are connected through HuBMAP's robust data provenance tracking (see below).

Spatial registration: Data submitters are obligated to submit spatial coordinate information about tissue blocks through the Registration User Interface (RUI), developed by the Human Reference Atlas (HRA) team[55]. The RUI captures "tissue extraction site[s] in relation to a 3D reference organ" using the Common Coordinate Framework (CCF)[33].

### *Data Ingestion*

Once a HuBMAP-funded component submits a bulk data upload via the HuBMAP ingest portal, HIVE data curators are notified via email and Slack automations that the data upload is ready for ingestion. Ingestion has five parts: validation, reorganization, processing, quality assurance and quality control (QA/QC), and publication. Validation for bulk uploads leverages automated tests for metadata file formatting and values using an API powered by the CEDAR Metadata Spreadsheet Validator. The validation workflow also checks individual dataset directories in a bulk upload for file presence and corruption (with particular plugin testing for imaging files like OME-TIFFs and sequencing files like FASTQ). Validation instructions and code can be found at https://github.com/hubmapconsortium/ingest-validation-tools and https://github.com/hubmapconsortium/ingest-validation-tests. After an upload is validated, data curators trigger automated reorganization of the bulk upload's data and metadata files from the Globus-managed staging directory into distinct Globus-managed dataset directories with unique HuBMAP IDs and UUIDs. Both validation and processing (see below) use the Apache Airflow open-source workflow management platform (https://airflow.apache.org). Once the reorganized datasets have been processed through data-type-specific uniform processing pipelines, data curators organize the primary (raw) and processed datasets for QA/QC by the relevant funded component. After approval for data publication is granted, the datasets are published, and DOIs are minted for them. After data publication, HuBMAP datasets are available for access by anyone from the portal, with the exception of protected raw human sequencing data, which is hosted and made available on dbGaP following access approval. Links to HuBMAP data in dbGaP are provided in context for all human sequencing datasets within the portal.

## HuBMAP Data Organization

### *Processing*

The HuBMAP Data Portal contains datasets from a wide variety of modalities and data collection methodologies, with some measurement types represented by multiple assay variants, e.g., measurement of gene expression via single-cell/single-nucleus RNA sequencing. Integration of data between these assay variants can be a significant practical challenge, one which the HuBMAP Consortium has addressed by adopting a single analysis pipeline for each high-level assay category, internally run on consortium data to ensure harmonized and comparable outputs. Usage of a uniform computational pipeline (accounting for assay differences as appropriate) allows for harmonization of data processing results between these assay variants, tissues, donors, and data providers.

HuBMAP computational pipelines are implemented as workflows in the Common Workflow Language (CWL)[56]. CWL workflows are composed of command-line tools with clearly defined input and output files, and in HuBMAP usage, these command-line tools are encapsulated in reproducibly buildable Docker images, allowing deterministic pipeline builds and executions. The HuBMAP pipeline release process consists of tagged releases of Git repositories and associated Docker images, ensuring consistent deployment and execution of pipeline releases in HuBMAP computational resources.

HuBMAP data processing composes multiple high-level CWL workflows into the analysis steps performed for datasets of each assay type. This workflow separation manages dispatch of distinct pipeline steps to CPU or GPU nodes as appropriate, in addition to execution of distinct data steps for data analysis, reorganization, and creation of visualization assets. Pipeline versions are shown on the portal view of each processed dataset page, including versions of the relevant computational workflows, allowing for reproducible execution.

Computational pipeline outputs are standardized between assays, using cross-platform and cross-language formats when possible, and prioritizing rich interoperable formats like HDF5 over plain text formats like CSV. The HuBMAP Consortium heavily uses the AnnData[57] format, which provides a clear separation between primary measurements (e.g., protein or gene expression) and additional data or metadata tables/matrices such as observation or sample metadata or spatial coordinates of a cell or other structure.

Processing pipelines are data-type-specific, with distinct analytical steps and outputs for bulk, single-cell, and spatial data types. Bulk data types undergo quantification only; bulk RNAseq data are processed using Salmon[58] for transcript quantification, and bulk ATACseq data are processed using MACS2[59,60] for chromatin accessibility quantification. Single-cell and single-nucleus transcriptomic data (sc/snRNAseq) and 10x Multiome datasets are processed using a Salmon-based quantification pipeline with downstream analysis via Scanpy[61], including unsupervised Leiden clustering[62]; for supported tissues (currently heart, lung, and kidney), automated cell type annotation is additionally applied via Azimuth[14]. Single-nucleus ATACseq data are processed using ArchR[63], and chromVAR[64] for chromatin accessibility quantification and clustering, without cell type annotation. Spatial proteomics datasets (CODEX/PhenoCycler, Cell DIVE, MIBI) undergo image preprocessing, cell and nuclear segmentation, and per-cell feature extraction and unsupervised clustering via Spatial Process and Relationship Modeling (SPRM)[65]; uniform cell type annotation is not currently applied through the pipeline. Spatial transcriptomics datasets (Visium, Slide-seq) are processed using the same Salmon-based quantification pipeline as sc/snRNAseq and go through a Squidpy[66] analysis in addition to Scanpy, with outputs provided as SpatialData[67] objects containing co-registered images and spatial coordinates alongside molecular measurements. Pipeline descriptions and open-source code are available at https://github.com/hubmapconsortium and in Börner et al.[5]. Processing information for any given HuBMAP dataset can be found under the Protocols & Workflow Details section on a dataset page. For datasets with cell or nuclear segmentations, details are included under the Segmentation Channels & Quality section directly below the dataset visualization.

### *Provenance*

Provenance information is collected for all data ingested in the HuBMAP Data Portal. This provenance data captures information about various entity types involved in data derivation, including donors, organs, tissue samples, and in some cases other datasets from which the ingested data was derived. Following the prov:Entity and prov:Activity constructs from the World Wide Web Consortium Provenance standard (W3C PROV-DM)[68], provenance is stored as a graph (Fig. 8), in which each entity is represented by a node, linked by activity nodes that record how each entity was generated.

The entity node types include Donor, Tissue Sample, and Dataset. Donors are at the top level of the HuBMAP provenance graph, and represent the human donors who donated the tissue from which resulting experimental data is derived. Tissue Samples are various types of tissues from the Donors with sub-types: 1) Organ: The organ from which tissue was obtained, always directly connected to a parent Donor in the provenance graph; 2) Tissue Block: A tissue block resected from an Organ; 3) Tissue Section: A section sliced from a Tissue Block; and 4) Suspension: A suspension of cells from a solid tissue (Organ or Block) or from bodily fluid. Lastly, Datasets represent registered data held by HuBMAP and are data derived from two sources: 1) Data generated by assays run against Tissue Samples; and 2) Data generated by processing other Dataset registered data. In some cases, a processed dataset is derived from multiple raw component datasets generated by distinct assays performed on the same tissue sample. For example, a SNARE-seq2 processed dataset combines separate RNAseq and ATACseq raw datasets. These multi-assay relationships are shown in the Dataset Relationship Diagram and Provenance Graph (Fig. 8).

Dataset versioning is supported on the portal. Versioned Datasets arise when processing pipelines are updated with additional functionality or newer versions of tools. Pipeline updates are applied selectively to relevant datasets, allowing the portal to incorporate improvements without disrupting the full dataset collection. Pipeline versions are tracked and visible on individual dataset pages, including versions of the relevant computational workflows. Users combining datasets processed at different points in time are advised to check pipeline versions to ensure consistent processing and compatibility for the intended analysis. An example of a versioned processed dataset is illustrated in Fig. 8. Versioning of assay-generated data is also supported, for instance, HBM937.RTLX.357 has been superseded by HBM588.GLJR.289.

## User-Driven Portal Design and Tool Functionality

A user-centric design approach was used to support exploration of complex biological human data on the HuBMAP Data Portal[6]. This design process was iteratively refined based on insights from user studies with domain experts, including semi-structured interviews, surveys, affinity mapping, and moderated usability testing sessions. These methods focused on understanding task-based workflows, identifying usability issues, and aligning functionality with real-world research needs. Resulting use cases (Table 1) span five categories: 1) Explore Biological and Single-Cell Data, 2) Explore Spatial and Tissue Context, 3) Analyze and Collaborate, 4) Access and Retrieve Data, and 5) Manage and Validate Data. These use cases cover a range of primary target users and scientific goals, which informed the design of associated portal tools and functionality.

In addition to structured user studies, we engaged directly with the community through consortium meetings, workshops combining live demos with open discussion to elicit use cases and feedback, hackathons, and conference roundtables and presentations. These in-person and virtual events provided opportunities to validate assumptions, gather requirements, and test existing features in realistic research workflows. An external UX consultant further conducted heuristic evaluations of the portal, led workshops to introduce user-centered design principles to internal stakeholders and provided expert feedback on design consistency, accessibility and interaction patterns. These community engagement activities validated our design approach and highlighted additional areas for refinement.

To complement these qualitative insights, we incorporated quantitative insights from portal usage analytics. Metrics such as pageviews, download statistics, and navigation patterns provided objective evidence of how interface updates influence user behaviors. Analytics also revealed areas where engagement lagged or tasks were frequently abandoned, helping to identify features requiring further refinement or clearer guidance. These metrics were reviewed monthly to monitor trends and assess the impact of design changes. An example of this was data downloads, where navigation data indicated frequent transitions between related raw and processed datasets, accompanied by low download completion rates and elevated exit rates. In response, we unified the information on a single page and introduced a bulk download feature to streamline access, enabling command-line transfers of multiple datasets at once.

Within the first year of this production phase, the portal also began supporting active users for tasks including data ingestion, processing, quality control, analysis, and publication. Features were expanded or iterated upon based on community feedback, active areas of consortium research, and emerging needs of the field such as new experimental technologies and the rapid increase in availability of 3D data. As data became increasingly available on the portal, the growing user need for integrated analyses drove the development of Workspaces. Workspace infrastructure was successfully used in two full-day hackathons, highlighting the need for a quick launch time, the ability to run multiple Workspaces at the same time, the ability to share Workspaces, template example pages, and improved addition of datasets to Workspaces—all of which have been implemented. Users have noted that the Workspaces are integrated well within the portal and work with large data, allowing users to easily test their workflows.

The design and development teams emphasized the importance of accessibility[69], which resulted in the implementation of a shortcut-based navigation tool[70] to support keyboard-only navigation of the site as well as a thorough review and remediation of accessibility issues with the help of a group of accessibility researchers. The colors in the portal user interface were optimized to be fully compliant with the Web Content Accessibility Guidelines (WCAG) AA standard[71].

Together, this mixed-methods user research strategy established a sustainable framework for user-centered development. By embedding UX best practices, combining qualitative and quantitative evidence and maintaining strong community engagement, we ensured that the portal continued to evolve with the needs of its users and the broader biomedical research community.

## Data Availability

All data presented in this paper are publicly available on the HuBMAP Data Portal, https://portal.hubmapconsortium.org. As the consortium is still publishing data, the volume of HuBMAP data is expected to continue to grow beyond the date of this publication. Dataset visualizations featured in Figure 5 are publicly available, with the exception of Xenium and GeoMx, which were undergoing QA at the time of publication. Interactive visualizations of datasets featured in Figure 5 are available on the HuBMAP Data Portal Publication Page: https://portal.hubmapconsortium.org/publications/hubmap-data-portal.

## Code Availability

All software used to construct the HuBMAP Data Portal infrastructure and UI is open source and freely available from repositories within the HuBMAP GitHub organization https://github.com/hubmapconsortium, from the Vitessce GitHub repository https://github.com/vitessce, and other repositories mentioned in the text. Documentation is available at https://docs.hubmapconsortium.org/.

## Acknowledgements

We acknowledge Chuck McCallum for leading an initial implementation of the portal and Jeremy Kriegel for UX consulting support. We also thank Sehi L'Yi and Lawrence Weru for providing accessibility expertise to the development team.

## Funding Statement

This research is supported by the NIH Common Fund, through the Office of Strategic Coordination/Office of the NIH Director under awards OT2 OD033758 (N.G.), OT2 OD033759 (P.D.B.), and OT2 OD033761 (M.M.R.).

## Author Contributions

The manuscript was written by M.L.T. with the help of T.C.S., T.S.L., B.H., B.S., L.C., M.S.K., D.L., R.M., M.H., M.M.R., P.D.B., and N.G. The software was developed by T.C.S., B.S., L.C., N.A., S.A., D.Be., K.B., I.C-B., J.C., C.C., P.C., S.D., S.F., D.F., J.H., T.K., M.S.K., D.L., X.L., A.Mo., E.M., J.M., M.J.O., G.P., S.S., M.S., A.S., X.T., J.W., Z.Y., M.H., M.M.R., P.D.B., and N.G. The portal UI/UX design was led by T.S.L. The data curation was handled by B.H., S.A., D.Bo., I.C-B., S.F., J.H., R.M., M.J.O., A.J.R., S.S., A.S., X.T., and J.W. The data visualizations were developed by M.L.T., T.C.S., T.S.L., L.C., N.A., J.C., T.K., M.S.K., A.Mo., E.M., and N.G. The computing resources and analysis tools were provisioned by C.C., E.H., R.S., D.S., and P.D.B. The project was managed and coordinated by M.L.T., N.G., P.D.B., J.S., M.M.R., M.H., S.F., B.S., L.C., Y.M., A.Mc., J.M., R.M., A.J.R., E.N., and M.A.M. The work was supervised by N.G., P.D.B., M.M.R., J.S., and M.H. with the support of M.L.T., B.S., L.C., S.F., M.A.M., R.S., E.H., A.J.R., and J.M.

# HuBMAP Consortium

**Members of the HuBMAP Consortium with direct authorship**

Morgan L. Turner[1], Thomas C. Smits[1], Tiffany S. Liaw[1], Brendan Honick[2], Bill Shirey[3], Lisa Choy[1], Nikolay Akhmetov[1], Shaokun An[4], David Betancur[2], Dominic Bordelon[2], Karl Burke[3], Ivan Cao-Berg[2], John Conroy[1], Chris Csonka[2], Penny Cuda[5], Sean Donahue[5], Stephen A Fisher[6], Derek Furst[3], Ed Hanna[2], Josef Hardi[7], Tabassum Kakar[1], Mark S. Keller[1], Devin Lange[1], Xiang Li[2], Yan Ma[1], Alison McWilliams[2], Austen Money[1], Richard Morgan[3], Eric Mörth[1], Juan Muerto[2], Mark A. Musen[7], Emily Nic[2], Martin J. O'Connor[7], Gesina Phillips[2], Alexander J. Ropelewski[2], Ryan Sablosky[2], Sravani Saripalli[4], Max Sibilla[3], Derek Simmel[2], Alan Simmons[3], Xu Tang[4], Joel Welling[2], Zhou Yuan[3], Martin Hemberg[4], Matthew Ruffalo[5], Jonathan Silverstein[3], Philip Blood[2], and Nils Gehlenborg[1]

**Members of the HuBMAP Consortium with no direct authorship**

Sammi Abate[8], Haitham Abdelazim[9], Roselkis M. Adames[1], Sierra M. Adkins[10], Jumana Afaghani[7], Abdalla Ahmed[7], Kyung Jin Ahn[11], Mimi Alkattan[1], Jamie Allen[12], Erick Alvarado[10], Antonella Arruda de Amaral[13], Pranay Ammula[4], David Anderson[12], Nate Anderson[14], Christopher R. Anderton[15], Michael Angelo[7], Ishaq Ansari[9], Charles Ansong[15,16], Siddharth Apte[14], Lillian Atchison[9], Demilade Awosika-Olumo[17], Rafaela Baggio[7], Amir Bahmani[7], Huajun Bai[11], Amanpreet K. Bains[18,19], Rachel Bajema[14], Karol Balderrama[20], George Bandek[1], Gautam Bandyopadhyay[21], Ziv Bar-Joseph[5], Laura Barisoni[22], Ross Barnowski[23], Daria Barwinska[14], Aidan Bauer[11], Laren Becker[7], Winston R. Becker[7], Gurjaspal Singh Bedi[14], Ken Bedi[6], Christina Beharry[9], Sean Bendall[7], Keith Bettinger[7], Joanna Y. Bi[7], Supriya Bidanta[14], Axel Bolin[14], Eric Boone[10], Gregory T. Booth[24], Samuel Border[9,25], Marc Bosse[7], William Bowen[14], Lisa M. Bramer[15], Maya Brewer[12], Christine Briggs[1], Maigan Brusko[9], Justin Buchanan[10], Andreas Bueckle[14], Christopher Daniel Burke[85], Kristin E. Burnum-Johnson[15], Andrew Butler[26], Kevin Matthew. Byrd[27,28,29], Katy Börner[14], Alexander Cabrera[6], Elicia Cabrera[11], Long Cai[23], Martha Campbell-Thompson[9], Dongfeng Cao[30], Richard Caprioli[12], Chiara Caraccio[7], Anita Caron[31], Megan Carroll[2], James P. Carson[17], Anelizze Castro-Martinez[10], Zimu Cen[11], Carrie Certo[3], Kexin Cha[7], Chrystal Chadwick[8], Thomas Chan[1], Siddhi Chavan[14], Angela Chen[13], Beishan Chen[11], Daiqing Chen[18], Fei Chen[20], Haoran Chen[5], Jie Chen[3], Jing Chen[9], Junchen Chen[10], Lu Chen[32], Xu Chen[33], Willi Cheung[10], Ken Chiacchia[2], Sanghee Cho[8], Geremy Clair[15], Wasserfall Clive[9], Madeline Colley[12], Kimberly Conklin[10], James M. Considine[18,34], Anthony Corbett[21], Alex Corwin[35], Alexis Costa[21], Daniel L. Cotter[7], Elise T. Courtois[36], Risa Cozza[10], Joseph Craft[37], Leonard Cross[14], Alexandra Cuaycal[9], Christine Curcio[30], Kali J. Dale[38], Nina Dar[23], Aarti Darji[1], Shakeel Ahamed Davanagere[14], Riza M. Daza[24], Mark deCaestecker[12], Stephen Deems[2,39], David J. Degnan[15], Gail Deutsch[24], Sumanth Devarasetty[9], Dinh Diep[10,40], Katerina Djambazova[12], Sergii Domanskyi[36], Durga Sritha Dongla[9], Qianhui Dou[41], Richard Drake[42], Satyam Dubey[14], Michael Duffy[6], Martin Dufresne[12], Thu Elizabeth. Duong[10], Jennifer Dutra[21], Michael T. Eadon[14], Quinn T. Easter[27], Kirk Ebelhar[14], Anushka Edlabadkar[10], Tarek M. El-Achkar[14], Mohamed Y. ElSadec[1], Diane Eshelman[43], Almudena Espin-Perez[7], Edward D. Esplin[7], Allison Esselman[12], Brent Ewing[24], Samuel L. Ewing[9], Louis Falo[3], Jean Fan[44], Rong Fan[37], Melissa Farrow[12], Patricia Favaro[7], Michael J. Ferkowicz[14], Erik Ferlanti[17], Ricardo Melo Ferreira[14,45], Shane Filus[2], Kathleen Fisch[10], Katelyn Flowers[20], William F.

Flynn[36], Agnes Fogo[12], Patrick Foley[13], Davide G. Franchina[7], Michael Gallant[14], Lu Gan[11], Ruli Gao[46], Ethan Gaskin[2], Kyle Gaulton[10], James Gaupp[12], James C. Gee[6], Soumya Ghose[8], Fiona Ginty[8], Debora Gisch[14], Madeleine Givant[13], Ilan Gold[1], Joana Goncalves[47], Xuan Gong[48], Ken Goodwin[2], Brittney Gorman[15], Rula Green-Gladen[24], William J. Greenleaf[7], Rudi Gunawan[25], Ziyuan Guo[49], Rajat Gupta[1], Shubham Gupta[7], Danielle Gutierrez[12], James S. Hagood[28], Abbas Hakim[10], Kyu Sang Han[44], Rongduo Han[51], Seung Yub Han[6], Kevin Hanshaw[52], Yuhan Hao[26], Raymond Harris[12], Mamon M. Hatmal[7], Yongqun Oliver He[53], Randy Heiland[14], Jesse Helfer[3], David Helminiak[54], Emerson Hernly[55], John W. Hickey[22], Annie Hiniker[33], Samantha A. Hoffman[9], Jeanne Holden-Wiltse[21], Annika Holmstrom[9], Young-Kwon Hong[13], Jody Hooper[7], Aaron M. Horning[7], Brooke Howitt[7], Hang Hu[55], Po Hu[11], Qiwen Hu[1], Tiffany Hua[11], Molly Huang[10], Heidie Huyck[21], Bruce Herr II[14], Srisharanya Injarapu[9], Mushfeqa Iqfath[55], Alexandra Irger[1], Olof Isberg[12], Taro Ishimori[28], Maxim Itkin[6], Valerio Izzi[56], Dana L. Jackson[24], Marni B. Jacobs[10], Daniel Jacobsen[10], Sanjay Jain[40], Yashvardhan Jain[14], Fulcher M. James[15], Niteesha Jangam[14], Matthew Jehrio[21], Gordon Z. Jiang[13], Lihua Jiang[7], Lixia Jiang[7], LiXue Jiang[55], Sarah Johnston[6], Marda Jorgensen[9], Yingnan Ju[14], Nikolaos Kalavros[13], Peter Kant[57], Dimitra Karagkouni[13], Thomas V. Karathanos[7], Arivarasan Karunamurthy[3], Heather R. Kates[9], Madhurima Kaushal[40], Selin Kavak[9], Neil L. Kelleher[46], Mishal Khan[9], Peter Kharchenko[1], Juhi Khare[14], Junhyong Kim[6], Jessica Kirwan[9], Jocelyn Y. Kishi[48], Reta Birhanu Kitata[15], Holly Klinesmith[59], Amanda L. Knoten[40], Yongxin Kong[14], Angela Kruse[12], Da Kuang[6], Suhas Katari Chaluva Kumar[9], Yash Kumar[14], Anshul Kundaje[7], Yumi Kwon[15], James A. Labyer[17], Asmita K. Lagwankar[40], Blue Lake[10,60], Bennet Landman[12], Roy Lardenoije[47], Monica M. Laronda[61], Julia Laskin[55], Zoltan Laszik[62], Ken Lau[12], Louise C. Laurent[10], Angeannie Lefevre[13], Logan A. Lewis[15], Chenran Li[32], Jiaging Li[12], Xiangtang Li[55], Zhi Li[10], Jennifer Li-Pook-Than[7], Yen-Chen Liao[15,63], Duanduan Lin[11], Shin Lin[24,64], Skykar Lin[26], Yiing Lin[40], Scott A. Lindsay[10], Chunjie Liu[11], Huiping Liu[46], Yang Liu[37], Rachel Liu-Galvin[9], Victoria Liwang[9], Angelo Rosso. Llantada[27], Marie Lott[11], Martin Lotz[65], Lisa Lowery[8], Edward Lu[14], Peiran Lu[11], Paean Luby[14], Nicholas Lucarelli[9], Shefali Luley[14], Selena Luo[1], Tianze Luo[3], Yuling Ma[13], Paul Macklin[14], Evan Macosko[20], Libby Maier[14], Brusko Maigan[9], Morad Malek[12], David Manthey[66], Trevor Manz[1], Kenneth Margulies[6], Atkinson Mark[9], Matthew Martindale[14], Jennifer K. Marx[1], Cayla N. Mason[10], Clayton E. Mathews[9], Bruno Matuck[67,27], Peter Maye[68], Chuck McCallum[1], Elizabeth McDonough[8], Hannah McDowell[61], James McLaughlin[31], Liam McLaughlin[40], Morgan Meads[10], Jeffrey Messinger[30], Sydney Meyer[1], Lukasz Migas[47], Brittany Minor[40], Ravi S. Misra[21], Roger Moens[47], Jeffrey Moffitt[69], Gesmira Molla[26], Emma Monte[7], Liya Mooradian[11], Felipe Moser[47], Megan Mulholland[10], Hrishikesh Mulkutkar[14], Hiroaki Murano[28], Deborah G. Murdock[11], Robert F. Murphy[5], Kay Métis[3], Alexandra Naba[18], Sheethal Umesh Nagalakshmi[13], Ernesto S. Nakayasu[15], Chanond Nasamran[10], Charles Nelson[6], Elizabeth Neumann[12], Stephanie Nevins[7], Hieu Nguyen[6], Tram Nguyen[1], Debra J. Nigra[2], Garry P. Nolan[7], Erik Nordgren[6], Diana R. O'Day[24], Kathleen O'Neill[6], Kenichi Okuda[28], Merissa Olmer[65], Jacqueline Olness[10], Tyler Ostrander[10], Naval Pandey[14], Minxing Pang[6], Priyadarshini Pantham[10], Mana Parast[10], Keyur Parekh[14], HJ Park[3], Nichole Parker[1], Sahil Patel[10], Ushma Patel[14], Nikolaos Patikas[4], Heath (Nathan) Patterson[12], Anindya S. Paul[9], Ljiljana Paša-Tolić[15], Liming Pei[11], Taojin Peng[10], Ting Peng[11], Luiza Perucci[10], Thai Pham[12], Mina Pichavant[7], Paul Piehowski[15], Nico Pierson[23], Ellie Pingry[12], Yered Pita-Juarez[13], Ariana Pitonzo[21], Raf Van de Plas[47], Sylvia K. Plevritis[7], Nongluk Plongthongkum[10], Athanasios Ploumakis[13], Prasanth Potluri[11], Alison Pouch[6],

Alvin Powers[12], Kanella Prodromidou[13], Gloria S. Pryhuber[21], Aleix Puig-Barbe[31], Jeffrey Purkerson[21], Danial Qaurooni[14], Wei-Jun Qian[15], Ling Qin[6], Catherine Qing[14], Ellen Quardokus[14], Andrea J. Radtke[70], Presha Rajbhandari[71], Rebecca Rakow-Penner[10], Ramalakshmi Ramasamy[36], Amber Ramesh[14], Kumar Rashmi[15], Jessica Ray[9], David F. Read[24], Christian Reardon-Lochbaum[69], Lisel (Elizabeth) Record[14], Sashank Reddy[44], Stephanie Reinert[40], Lauraine Rivier[28], Paul Robson[36], Ana E. Rodríguez-Soto[10], Jean Rosario[6], Rachel Rose[35], David Rowe[68], Erin E. Rowell[61], Cihan Ruan[72], Nancy Ruschman[14], Angela Sabo[14], Sinem K. Saka[73], Pinaki Sarder[9], Sourav Sarkar[26], Rahul Satija[26], Diane Saunders[12,61], Shamit Gajanan Savant[9], Riley Sawka[11], Jacob Scherba[11], Leah Scherschel[14], Kevin Schey[12], Heidi Schlehlein[14], David Scholten[46], Robin Scibek[2], Ayellet V. Segrè[1], Maya Selvaraj[10], Kelly S. Shane[15], Siddharth Sharma[40], Shashank Shekhar[14], Jay Shendure[24], Michael Shewaregara[1], Lingyan Shi[10], Tujin Shi[15], Dong-Guk Shin[68], Rohan Shrestha[10], Gagandeep Singh[14], Dhruv Singhal[13], Santhosh Sivajothi[36], Anna Smith[12], Michael P. Snyder[7], Francesca Soncin[10], Anup Sood[8], Jeffrey Spraggins[12], Srimeenakshi Srinivasan[10], Sanjay Srivatsan[24], Valentina Stanley[10], Brent R. Stockwell[71], Yulia A. Levites Strekalova[9], Liwen Su[11], Shankar Subramaniam[10], Rasheed Sule[11], Xin Sun[10], Joel Sunshine[44], Christine Surrette[8], Vidyani Suryadevara[7], Hannah Swahn[65], Medina Sydykanova[14], Chi-Yi Tai[10], Kai Tan[11], Yuqi Tan[7], Sarah A. Teichmann[74,75,76,77], George Tellides[37], Anusha Thadi[11], Anusha Thaniana[11], Hua Tian[71], Leonoor Tideman[47], Brusko Todd[9], John E. Tomaszewski[25], Samson Toor[1], Katerina J. Trachtova[40], Cole Trapnell[24], Albert G. Tsai[7], Chia-Feng Tsai[15], Leo L. Tsai[41], Elizabeth Tsui[61], Tina Tsui[12], Daisy Unsihuay[55], Jackie Uranic[2], Mereena Ushakumary[15], Shikhar Uttam[3], David Van Valen[23], Heidi Vandyk[15], Saanvi Vavilala[11], Dušan Veličković[15], Marija Veličković[15], Margaret Vella[1], Jamie Verheyden[10], Jorge Villazon[10], Ioannis S. Vlachos[13], Jessica Waldrip[3], Nazifa Wali[55], Douglas C. Wallace[11], Allen Wang[10], Fusheng Wang[32], Mei-Yu Wang[2], Meng Wang[53], Shuoshuo Wang[13], Xin Wang[1], Xuefei Wang[23], Andrew Watts[21], Griffin Weber[1], Bei Wei[78], Annika K. Weimer[7], Gabrielle A. Widjaja[11], Eric Williams[8], Sarah M. Williams[15], Seth Winfree[79,80], Denis Wirtz[44], Devin Wright[14], Pei-Hsun Wu[44], Yi Xing[11], Tianyang Xu[10], Haichun Yang[12], Joan Yang[7], Manxi Yang[55], Li Yao[81], Dong Hye Ye[82], Peng Yin[48], Ruichuan Yin[55], Yin Yin[13], Kevin Yoo[1], Joseph Yracheta[38], Wenbao Yu[11], Jina Yun[23], Ali Zahraei[12], Kevin J. Zemaitis[15], Bo Zhang[40], Kun Zhang[10,60], Shiping Zhang[11], Ted Zhang[5], Zhengyan Zhang[40], Bingqing Zhao[7], Juanjuan Zhao[11], Zoey Zhao[10], Sheng Zhong[10], Mowei Zhou[15,83], Bokai Zhu[7], Chenchen Zhu[7], Quan Zhu[10], and Ying Zhu[15,84]

[1]Harvard Medical School, Boston, MA, USA. [2]Pittsburgh Supercomputing Center, Carnegie Mellon University, Pittsburgh, PA, USA. [3]University of Pittsburgh, Pittsburgh, PA, USA. [4]Gene Lay Institute of Immunology and Inflammation, Brigham and Women's Hospital, Harvard Medical School and Massachusetts General Hospital, Boston, MA, USA. [5]Carnegie Mellon University, Pittsburgh, PA, USA. [6]University of Pennsylvania, Philadelphia, PA, USA. [7]Stanford University, Stanford, CA, USA. [8]GE HealthCare. [9]University of Florida, Gainesville, FL, USA. [10]University of California San Diego, San Diego, CA, USA. [11]Children's Hospital of Philadelphia, Philadelphia, PA, USA. [12]Vanderbilt University, Nashville, TN, USA. [13]Beth Israel Deaconess Medical Center, Boston, MA, USA. [14]Indiana University, Bloomington, IN, USA. [15]Pacific Northwest National Laboratory, Richland, WA, USA. [16]National Institute of General Medical Sciences (NIGMS), Bethesda, MD, USA. [17]The University of Texas at Austin, Austin, TX, USA. [18]University of Illinois Chicago, Chicago, IL, USA. [19]University of Leipzig, Leipzig, Germany. [20]Broad Institute of MIT and Harvard, Cambridge, MA, USA. [21]University of Rochester Medical Center, Rochester, NY, USA. [22]Duke University, Durham, NC, USA. [23]California Institute of Technology, Pasadena, CA, USA. [24]University of Washington, Seattle, WA, USA. [25]University at Buffalo, Buffalo, NY, USA. [26]New York

Genome Center, New York, NY, USA. [27]Department of Oral and Craniofacial Molecular Biology, Philips Institute for Oral Health Research, Virginia Commonwealth University, Richmond, VA, USA. [28]Division of Oral and Craniofacial Health Sciences, Adams School of Dentistry, University of North Carolina at Chapel Hill, Chapel Hill, NC, USA. [29]VCU Massey Comprehensive Cancer Center, Virginia Commonwealth University, Richmond, VA, USA. [30]University of Alabama at Birmingham, Birmingham, AL, USA. [31]EMBL-EBI, Hinxton, Cambridge, UK. [32]Stony Brook University, Stony Brook, NY, USA. [33]University of Southern California, Los Angeles, CA, USA. [34]AbbVie. [35]GE Aerospace. [36]The Jackson Laboratory, Bar Harbor, ME, USA. [37]Yale University, New Haven, CT, USA. [38]Native BioData Consortium, Eagle Butte, SD, USA. [39]Internet2. [40]Washington University in St. Louis, St. Louis, MO, USA. [41]Mass General Brigham, Boston, MA, USA. [42]Medical University of South Carolina, Charleston, SC, USA. [43]Donnelley Financial Solutions (DFIN). [44]Johns Hopkins University, Baltimore, MD, USA. [45]Indiana University Indianapolis, Indianapolis, IN, USA. [46]Northwestern University, Evanston, IL, USA. [47]Delft University of Technology, Delft, Netherlands. [48]Harvard University, Cambridge, MA, USA. [49]Cincinnati Children's Hospital Medical Center, Cincinnati, OH, USA. [50]University of North Carolina at Chapel Hill, Chapel Hill, NC, USA. [51]Nankai University, Tianjin, China. [52]Google. [53]University of Michigan, Ann Arbor, MI, USA. [54]Marquette University, Milwaukee, WI, USA. [55]Purdue University, West Lafayette, IN, USA. [56]University of Oulu, Oulu, Finland. [57]Otsuka Precision Health. [58]Wyss Institute for Biologically Inspired Engineering, Harvard University, Boston, MA, USA. [59]Thermo Fisher Scientific. [60]Altos Labs. [61]Ann & Robert H. Lurie Children's Hospital of Chicago, Chicago, IL, USA. [62]University of California San Francisco, San Francisco, CA, USA. [63]Department of Translational Neurosciences, College of Medicine, University of Arizona, Phoenix, AZ, USA. [64]Emory University, Atlanta, GA, USA. [65]Scripps Research, La Jolla, CA, USA. [66]Kitware. [67]Hospital Israelita Albert Einstein, São Paulo, Brazil. [68]University of Connecticut, Storrs, CT, USA. [69]Boston Children's Hospital, Boston, MA, USA. [70]National Institutes of Health, Bethesda, MD, USA. [71]Columbia University, New York, NY, USA. [72]Santa Clara University, Santa Clara, CA, USA. [73]Genome Biology Unit, European Molecular Biology Laboratory (EMBL), Heidelberg, Germany. [74]Wellcome Sanger Institute, Wellcome Genome Campus, Hinxton, Cambridge, UK. [75]Cambridge Stem Cell Institute, Jeffrey Cheah Biomedical Centre, University of Cambridge, Cambridge, UK. [76]Department of Medicine, University of Cambridge, Cambridge, UK. [77]CIFAR, Toronto, ON, Canada. [78]Sun Yat-sen University, Guangzhou, China. [79]University of Nebraska Medical Center, Omaha, NE, USA. [80]Quantitative Cell Diagnostix. [81]Liyaovisuals Design Studio. [82]Georgia State University, Atlanta, GA, USA. [83]Department of Chemistry, Zhejiang University, Hangzhou, Zhejiang, China. [84]Department of Microchemistry, Proteomics, Lipidomics and Next Generation Sequencing, Genentech. [85]Department of Biomedical Informatics, University of Pittsburgh, Pittsburgh, PA, USA.

## Competing Interests

N.G. is a co-founder and equity owner of Datavisyn. A.J.R. has equity interest in ILMN and TXG. All other authors declare no competing interests.

| Use Case | Primary Target Users | Scientific Goal | Portal Tools | Impact |
|---|---|---|---|---|
| **Explore biological and single-cell data** | | | | |
| Explore available HuBMAP data | All users | Browse and filter public HuBMAP datasets by organ, assay, donor metadata, and more to identify data of interest | Datasets Search (Filter & Browse) | Enable any user to orient to available data and identify starting points for biological discovery |
| Explore datasets via natural language | Experimental biologists, clinical researchers, students | Retrieve specific datasets matching targeted biological criteria through conversational query | Datasets Search (Say & See) | Lower barrier to data discovery for users with limited computational expertise or familiarity with structured search interfaces |
| Explore cell types and gene expression across tissues | Experimental biologists, translational scientists, computational biologists, clinical researchers | Identify cell types across organs, view abundance and tissue location; examine variation in gene expression across tissues | Cell Type Pages, Biomarker & Cell Type Search, Cell Population Plot tool *scellop*, Organ Pages, Biomarker Pages | Enable targeted discovery of cell type distribution and tissue-specific gene expression to support biological discovery and annotation |
| Explore population-level cellular composition across demographics | Experimental biologists, computational biologists, clinical researchers | Investigate variation in cell types by age, sex, race or other metadata | Cell Population Plot tool *scellop*, Workspaces | Reveal biological variation across donor demographics to contextualize findings |
| Identify healthy tissue reference baselines for disease comparison | Translational scientists, clinical researchers | Investigate baseline molecular and cellular profiles in healthy tissue to support comparative or clinical studies | Datasets Search, Biomarker & Cell Type Search, Cell Type Pages, Organ Pages, Workspaces | Provide population-level reference ranges and healthy baselines that contextualize disease states and support diagnostic or therapeutic development |
| **Explore spatial and tissue context** | | | | |
| Explore 2D and 3D spatial organization of healthy tissue | Experimental biologists, computational biologists, clinical researchers | View spatial organization, morphology, and volumetric relationships of cells and anatomical structures in healthy human tissue, from whole organ to subcellular scale | Spatial Datasets, Vitessce, Vitessce Link, 3D Tissue Maps, Workspaces | Provide spatial and structural context for cell localization, tissue organization, and multiscale atlas exploration |
| Explore cellular neighborhoods and spatial interactions | Experimental biologists, computational biologists | Analyze cellular interactions, spatial co-localization patterns, and functional tissue units in context | Vitessce, SpatialQuery, Workspaces | Enable identification of cellular neighborhoods and spatial features |
| Explore | Experimental biologists, | Integrate and compare multiple | Multimodal Datasets, | Facilitate cross-modality |

| multimodal data within tissue context | computational biologists | omics modalities within tissue context | Workspaces | exploration to uncover complex tissue relationships |
|---|---|---|---|---|
| **Analyze and collaborate** | | | | |
| Analyze and visualize data in the browser | Experimental biologists, computational biologists, educators, students | Interact with large datasets via built-in tools and execute pre-configured workflows in a cloud-based environment without local setup | Workspaces, Vitessce, Datasets Search (Say & See) | Reduce computational barriers and support exploratory analysis |
| Organize and share datasets | All users | Curate, bookmark, and share datasets for projects, publications, or teaching | My Lists, Workspaces, Publication Pages, Collections, Vitessce | Enhance collaboration and reproducible dataset citations |
| **Access and retrieve data** | | | | |
| Download raw and processed datasets | Experimental biologists, translational scientists, computational biologists | Access and download standardized raw and processed datasets for reuse and validation | Globus and dbGaP, HuBMAP CLT | Enable reproducible results and integration into external workflows |
| Access metadata and data programmatically | Computational biologists, technologists | Automate metadata retrieval and integrate portal context into external pipelines | Public APIs for datasets and metadata (Search API, UBKG API) | Promote reuse and interoperability with external tools |
| Train AI models and benchmark computational tools | Computational biologists, data scientists | Identify large uniformly processed single-cell and spatial datasets for foundation model training, algorithm development, or tool validation | Datasets Search (Filter & Browse), Biomarker & Cell Type Search, Globus and dbGaP, HuBMAP CLT, public APIs | Enable development and validation of AI/ML models using standardized healthy human tissue data |
| **Manage and validate data** | | | | |
| Inspect and validate datasets prior to publication | Data submitters, consortium members | Review uniformly processed datasets for quality, apply domain-specific quality criteria, and approve data for publication | Vitessce (QA during ingestion), Dataset Relationship Diagram, Provenance Table and Graph | Ensure data quality and accuracy before public release; support transparent reproducible data provenance |

**Table 1: HuBMAP Data Portal use cases.** Primary target users, scientific goals, portal tools, and impact across five use case categories derived from user studies.

# Figure Legends

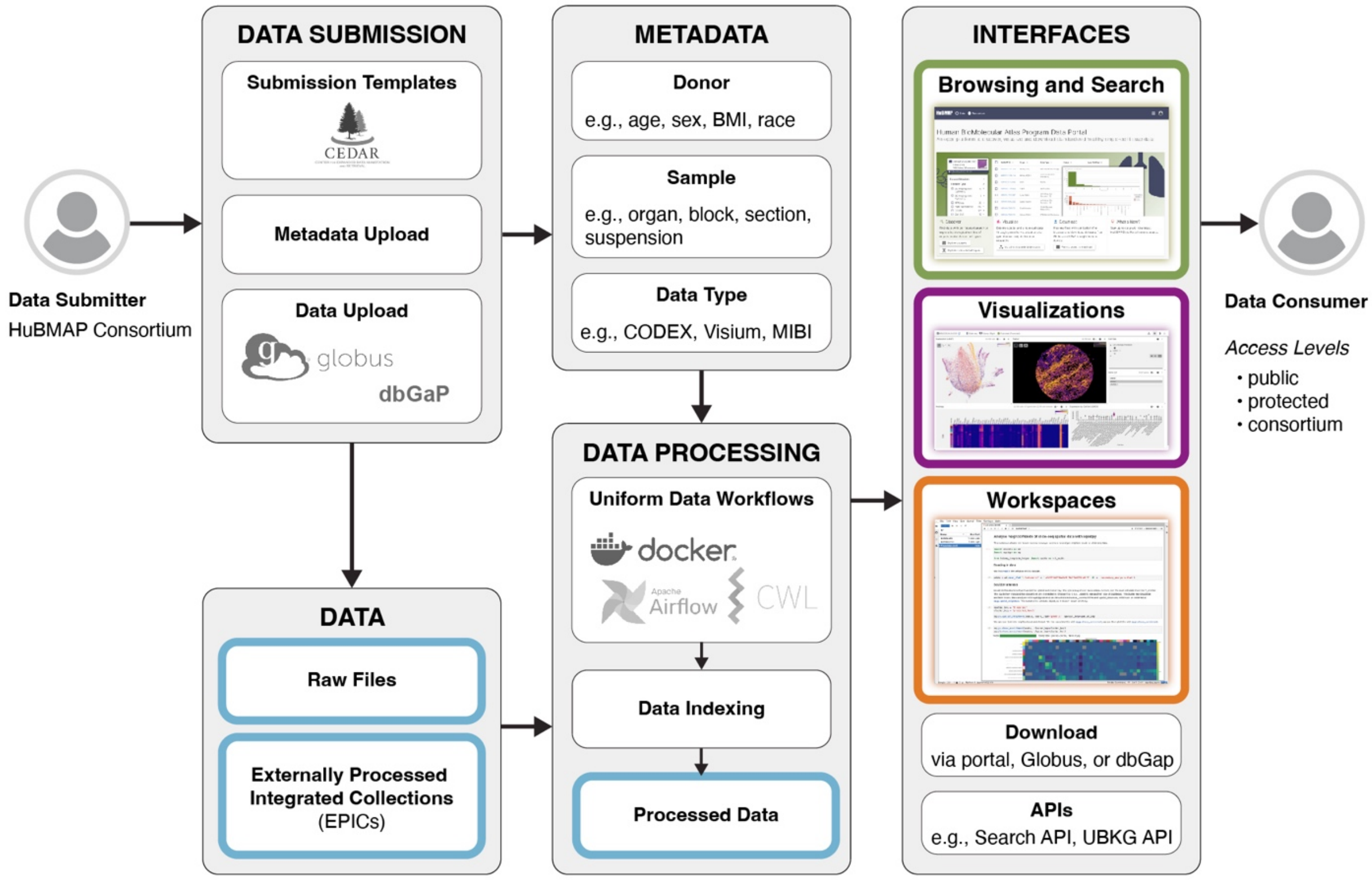

**Figure 1: Flow of data through HuBMAP Data Portal architecture.** Data submitters upload metadata and data via submission templates, which initializes ingestion and results in structured metadata and raw data files. Automated uniform processing pipelines produce derived outputs (e.g., processed data). Externally-processed data are submitted separately as Externally Processed Integrated Collections (EPICs). All resulting resources share harmonized data structures, enabling portal interfaces to uniformly operate across a diversity of data types. Within Portal user interfaces, this harmonized data (blue) can be searched and discovered (green), visualized (purple), and analyzed via Workspace notebooks (orange). This color coding is used consistently for these features in subsequent figures. Files are downloadable via the portal, Globus, or dbGaP; metadata is programmatically accessible via public APIs. Access is tiered: external users access published public data; authenticated users may access non-public datasets.

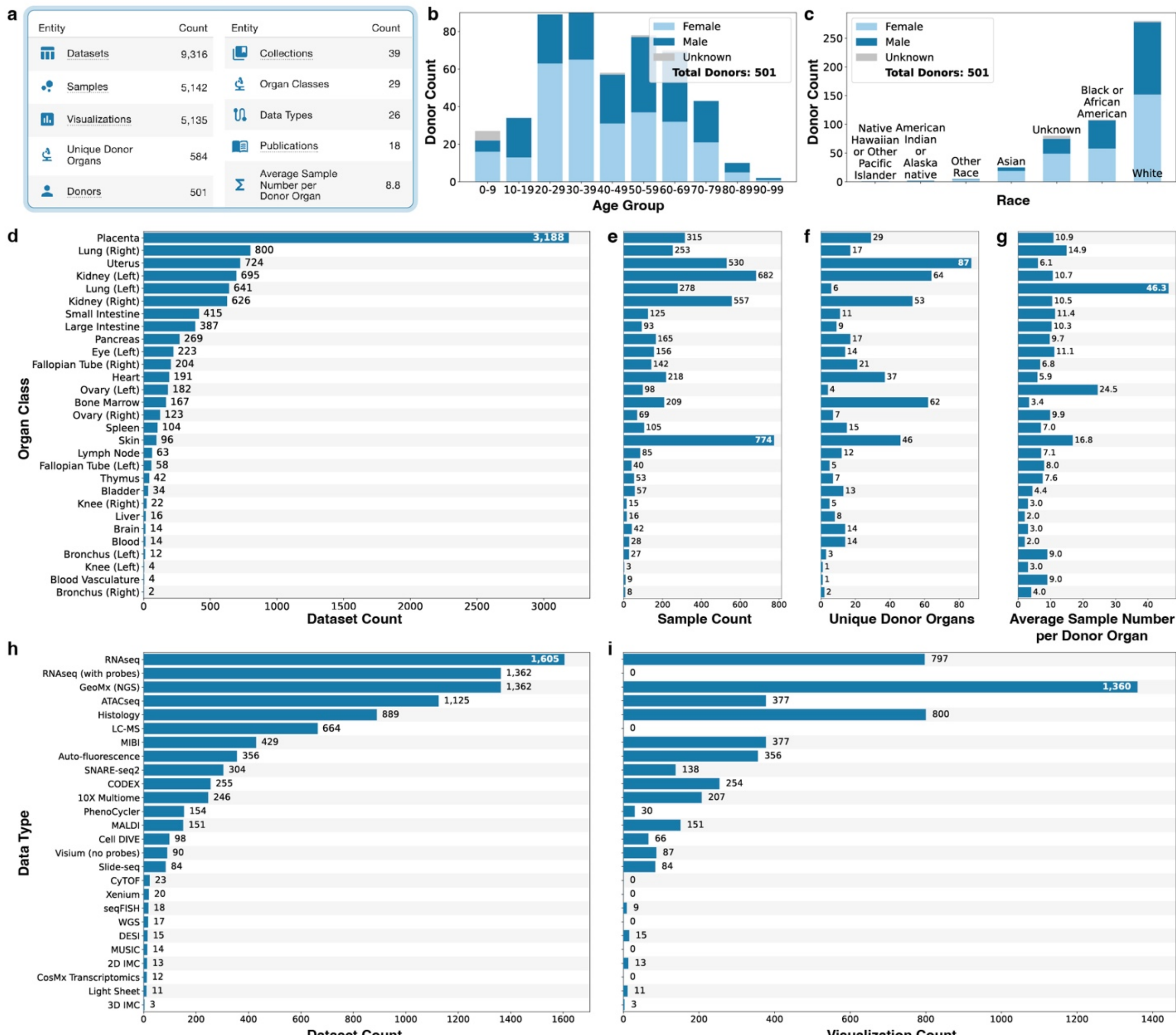

**Figure 2: Overview of data in the HuBMAP Data Portal. a** Summary of HuBMAP data entity counts. **b-c** Female (light blue) and male (dark blue) donor counts by (**b**) age group and (**c**) race; see Methods for details on donor metadata and use of "unknown". **d-g** Organ class distributions by (**d**) dataset count, (**e**) sample count, (**f**) unique donor organ count, and (**g**) average sample number per donor organ. **h-i** Data Type distributions by (**h**) dataset count and (**i**) Vitessce visualization count. Unique Donor Organs in (**f**) counts in distinct donor–organ pairs; a donor contributing samples across N organs is counted N times, so the bar sum exceeds the total unique donor count in panels (**b**) and (**c**). Data as of August 2026. Current data can be viewed at (https://portal.hubmapconsortium.org/data-overview).

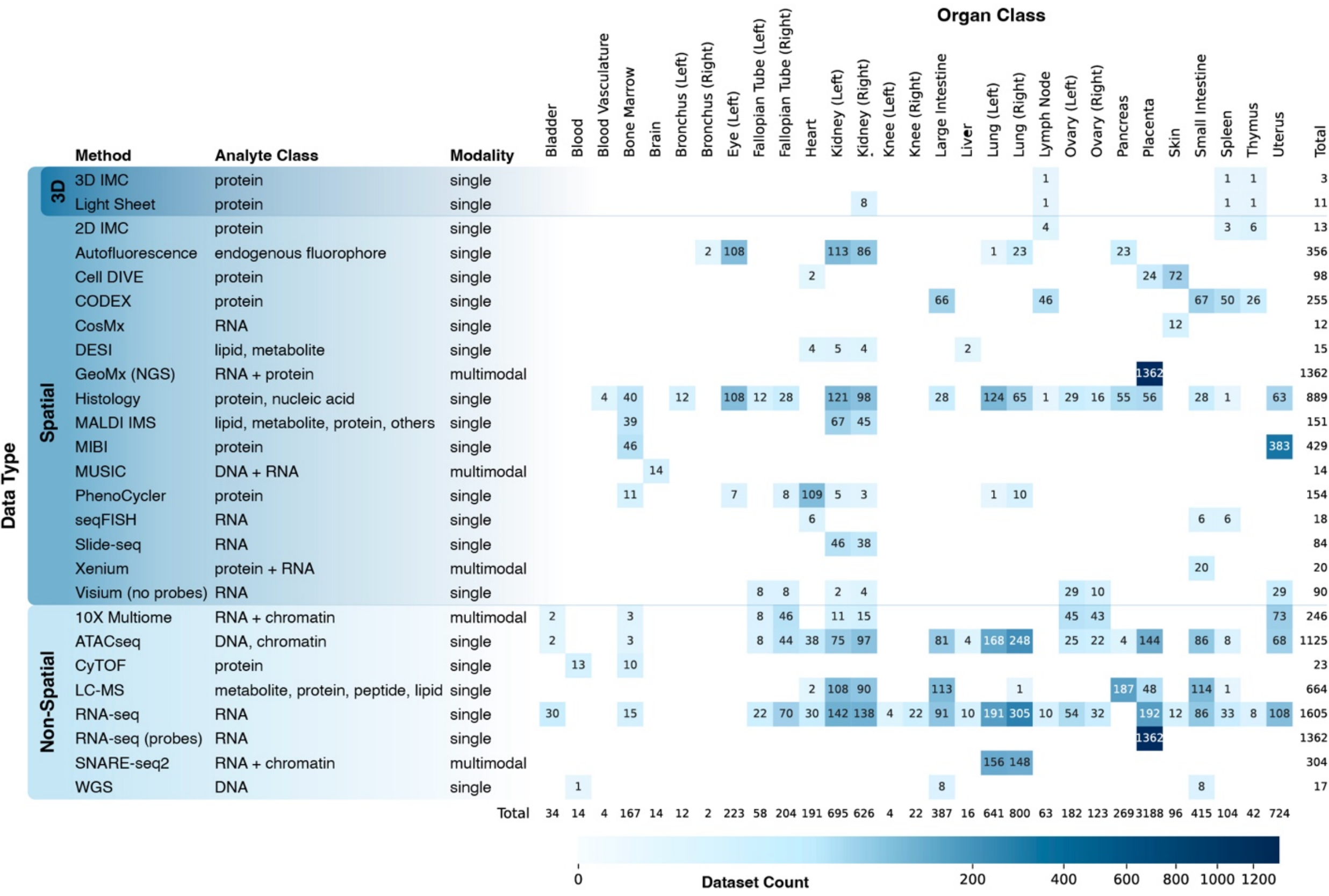

**Figure 3: Organs and data types on the HuBMAP Data Portal.** A table of data types and attributes aligned with a matrix of dataset counts per organ. Dataset counts are shaded on a color scale ranging from light blue (fewer datasets) to dark blue (more datasets). *RNAseq* and *ATACseq* each encompass bulk, single-cell, and single-nucleus variants. Acronyms: *ATACseq*, assay for transposase-accessible chromatin with high throughput sequencing; *CODEX*, co-detection by indexing; *CyTOF*, cytometry by time of flight; *DESI*, desorption electrospray ionization; *IMC*, imaging mass cytometry; *LC-MS*, liquid chromatography-mass spectrometry; *MALDI IMS*, matrix-assisted laser desorption ionization; *MIBI*, multiplex ion beam imaging; *MUSIC*, multinucleic acid interaction mapping in single cells; *RNAseq*, RNA sequencing; *seqFISH*, sequential fluorescence in situ hybridization; *SNARE-seq2*, single-nucleus chromatin accessibility and mRNA expression sequencing; *WGS*, whole-genome sequencing. Cell DIVE, CosMx, GeoMx, PhenoCycler, Slide-seq, 10x Multiome, Visium, and Xenium are technology or platform names, not acronyms.

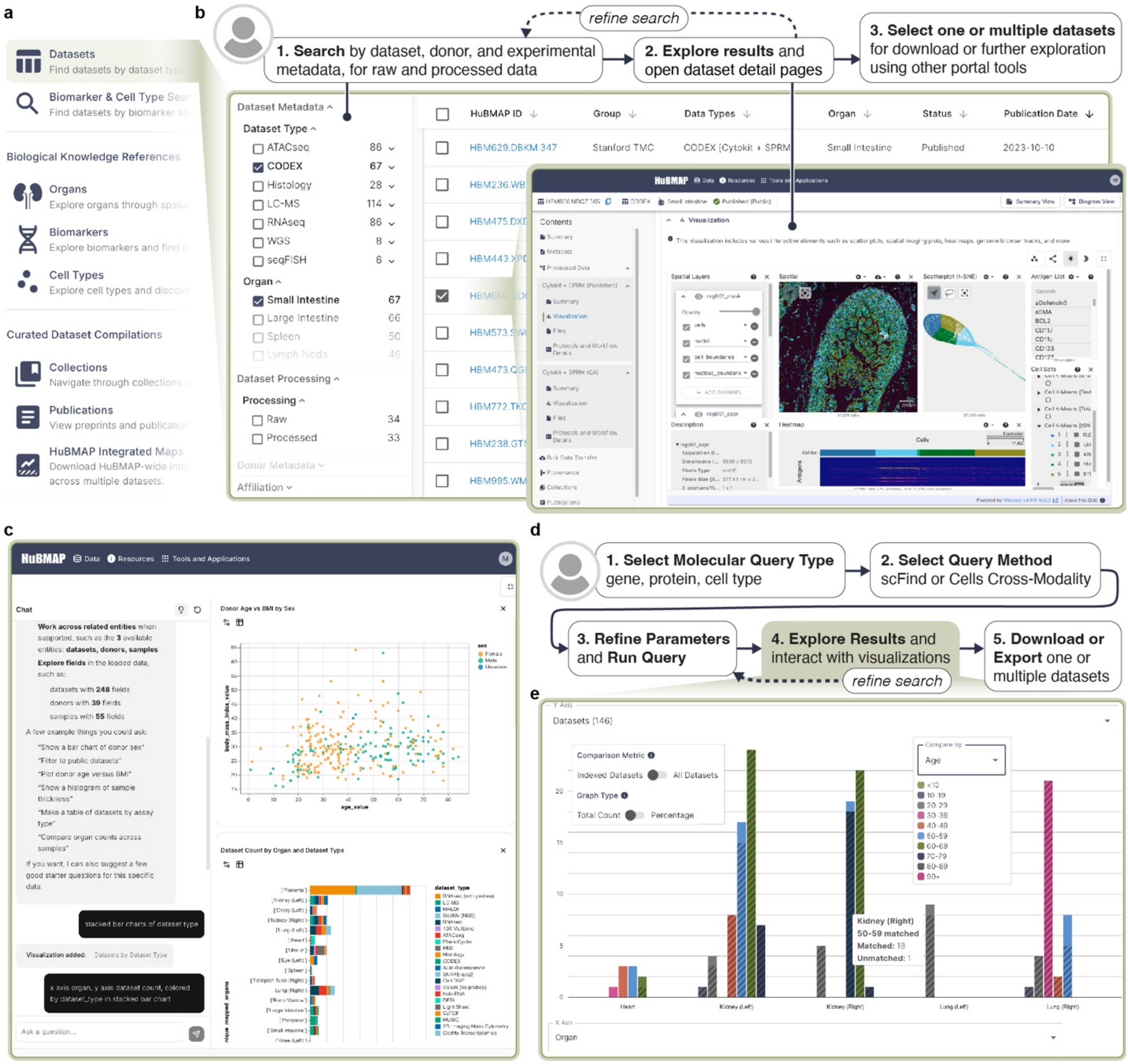

**Figure 4: Query user interfaces on the HuBMAP Data Portal. a** The Data menu provides access to HuBMAP data through curated dataset collections, metadata-driven queries, and molecular and cell-type queries. **b** The Filter & Browse mode of Dataset Search allows users to query dataset, donor, and experimental metadata to identify and explore datasets of interest. Datasets can be selected for further exploration or download; individual dataset pages display metadata, processed data, visualizations, provenance, attribution, and other associated information. **c** Say & See enables visualization-based data exploration through a natural language query interface. **d-e** The Biomarker & Cell Type Search allows users to search genes, proteins, and cell types. **d** User workflow. **e** Bar chart summarizing the results of a Cell-Type query for B Cells, with organ and donor metadata filters. Solid arrows represent user workflows connecting starting and ending points; dashed arrows represent optional refinement steps.

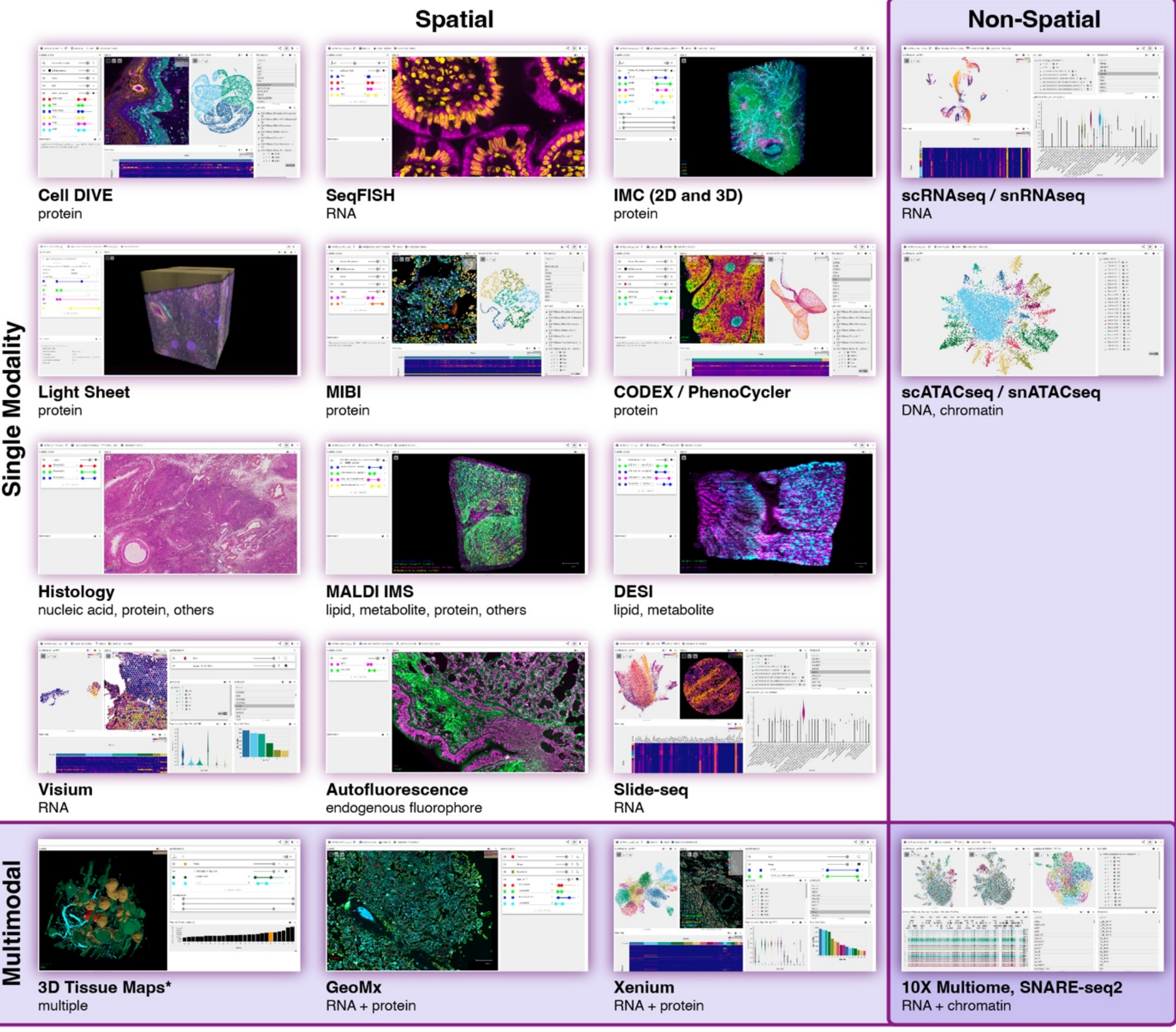

**Figure 5: Vitessce visualization diversity on the HuBMAP Data Portal.** The diversity of data types supported by Vitessce is organized in a matrix by spatial vs. non-spatial and single vs. multimodal. Unique visualization configurations are tailored to the analytes and spatial context of a specific data type. Acronyms: *ATACseq*, assay for transposase-accessible chromatin with high throughput sequencing; *CODEX*, co-detection by indexing; *DESI*, desorption electrospray ionization; *IMC*, imaging mass cytometry; *MALDI IMS*, matrix-assisted laser desorption ionization; *MIBI*, multiplex ion beam imaging; *RNAseq*, RNA sequencing; *seqFISH*, sequential fluorescence in situ hybridization; and *SNARE-seq2*, single-nucleus chromatin accessibility and mRNA expression sequencing. Cell DIVE, GeoMx, PhenoCycler, Slide-seq, 10x Multiome, Visium, and Xenium are technology or platform names, not acronyms. *3D Tissue Maps is a data format for multimodal volumetric data. Interactive visualizations of datasets featured in this figure are available on the HuBMAP Data Portal Publication Page: https://portal.hubmapconsortium.org/publications/hubmap-data-portal.

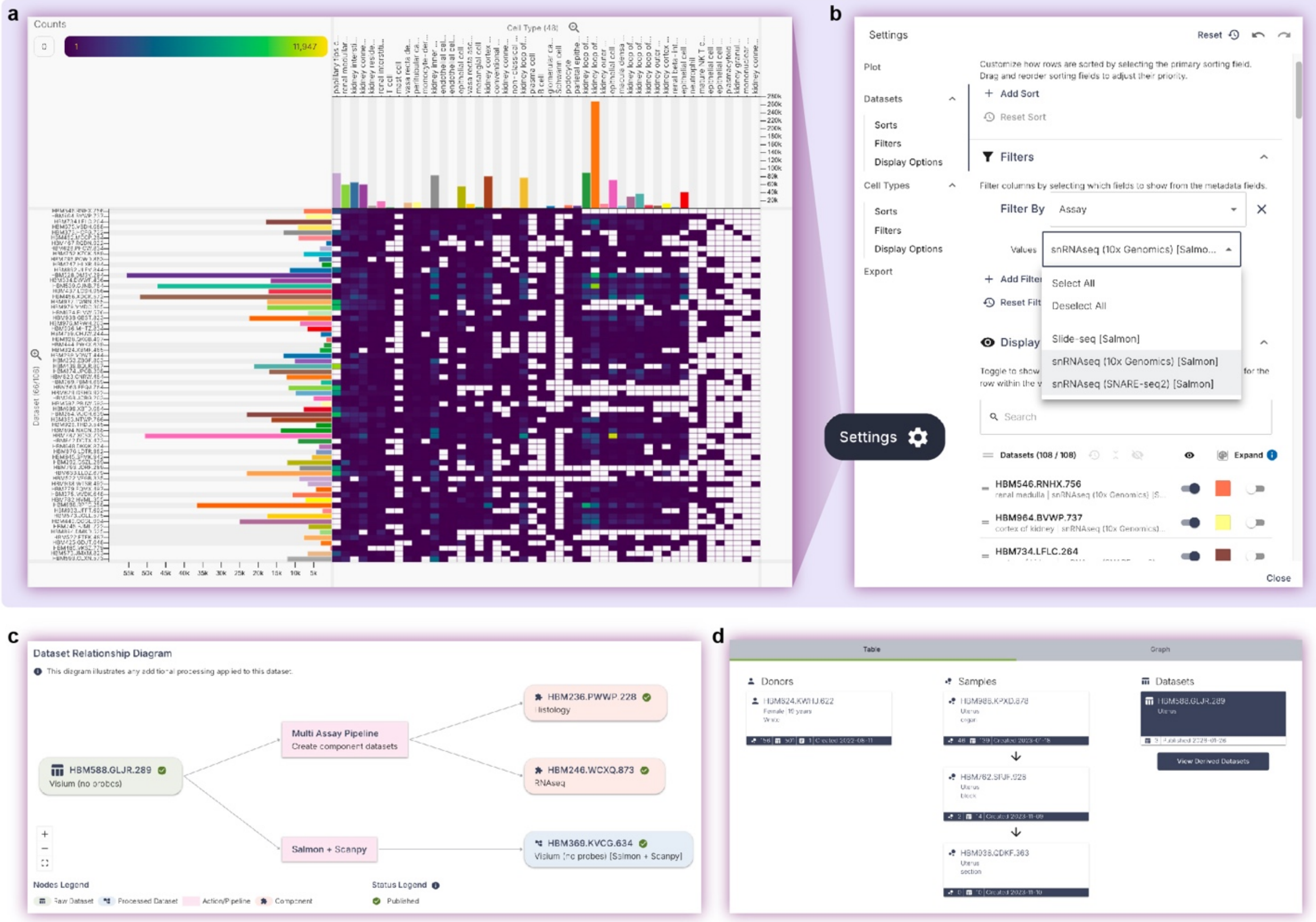

**Figure 6. Visualizations embedded across the HuBMAP Data Portal. a-b** Cell Population Plot viewer *scellop* on the Kidney Organ Page, showing, (**a**) cell type annotations across RNAseq kidney datasets and (**b)** the settings menu for configuration options. Interactive visualization: https://portal.hubmapconsortium.org/organ/kidney. **c** Dataset Relationship Diagram and **d** Provenance Diagram supporting provenance visualization on dataset detail pages.

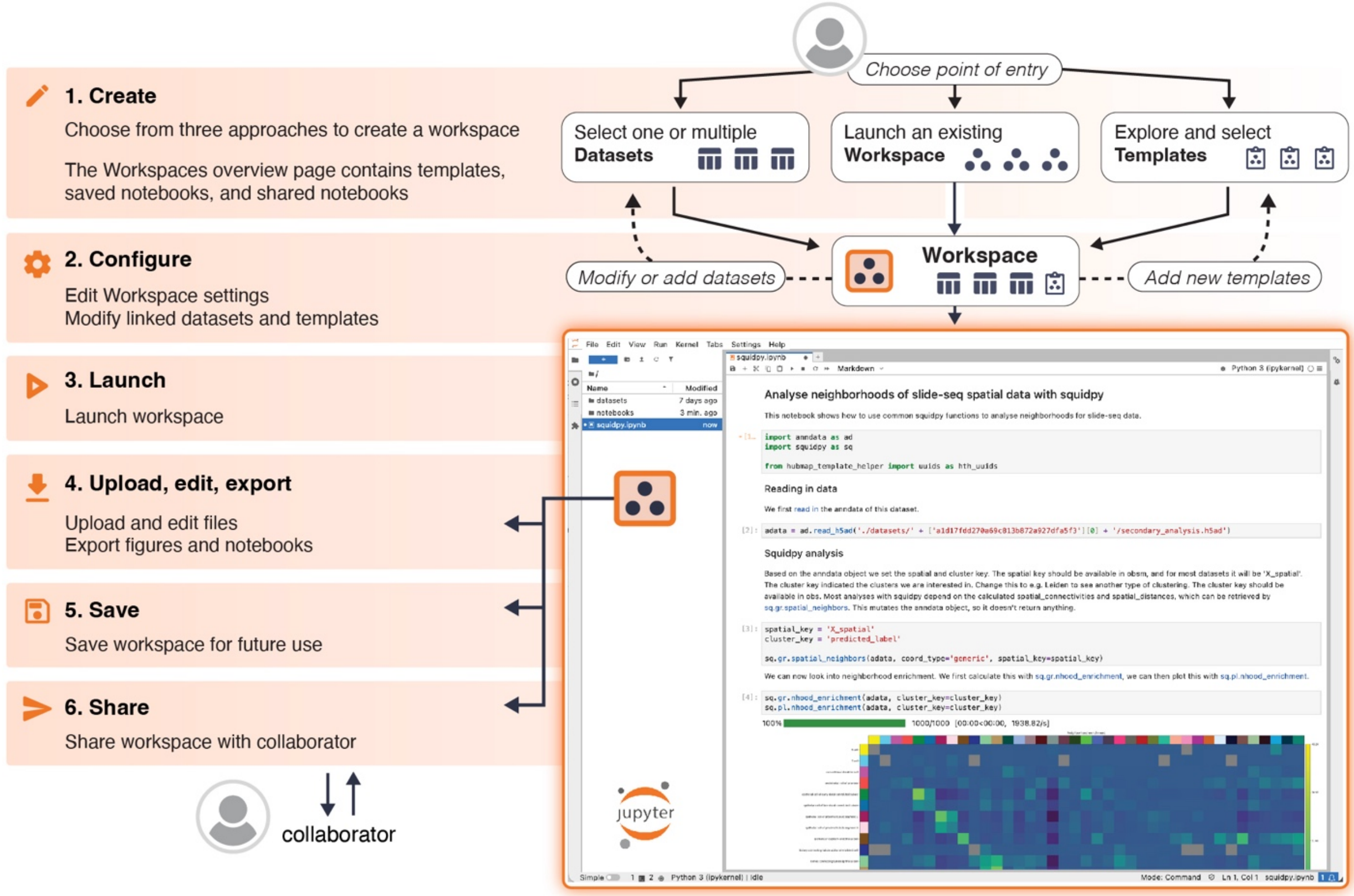

**Figure 7: Workspaces on the HuBMAP Data Portal.** Multiple entry points support workspace creation, exploration, collaboration, and export, enabling interactive analysis of HuBMAP data via Jupyter Notebooks in Python or R. Solid arrows represent user workflows connecting different starting and ending points; dashed arrows represent optional refinement steps.

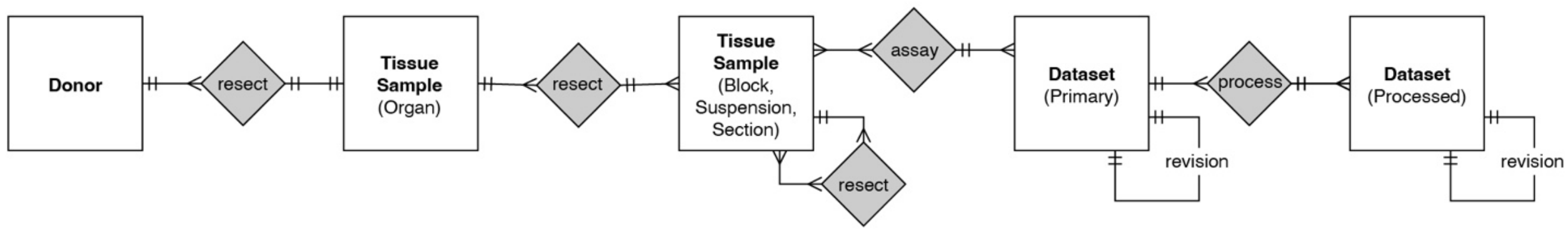

**Figure 8: Provenance of HuBMAP.** The HuBMAP Provenance model is a graph with entities (nodes) linked by activity nodes recording how each was generated, following W3C PROV-DM. Double lines indicate one-and-only-one relationships; split lines indicate many. Versioned datasets are shown in relation to their primary and processed datasets.